%% file: main.tex
\documentclass[letterpaper, 10 pt, conference]{ieeeconf}

\IEEEoverridecommandlockouts

\usepackage{graphics}
\usepackage{epsfig}
\usepackage{times}
\usepackage{hyperref}

\usepackage{amsmath,amsthm,amssymb}
\usepackage{mathtools}
\usepackage{mathrsfs}
\usepackage{subfiles}
\usepackage{enumerate}
\usepackage{color}
\usepackage[caption=false]{subfig}
\usepackage{balance}

\newtheorem{theorem}{Theorem}

\newtheorem{proposition}[theorem]{Proposition}
\newtheorem{definition}{Definition}
\newtheorem{remark}[definition]{Remark}

\newcommand{\R}{\mathbb{R}}
 \newcommand{\Rplus}{\R_{\geq 0}}

\newcommand{\C}{\mathcal{C}}

\newcommand{\classK}{\mathcal{K}}
\newcommand{\classKinfty}{\mathcal{K}_{\infty}}

\newcommand{\gap}{\vspace{.1cm}}

\newcommand{\newsec}[1]{\gap {\bf \noindent #1 }}

\newcommand{\Kclf}{K_{\mathrm{clf}}}
\newcommand{\Kcbf}{K_{\mathrm{cbf}}}

\title{\LARGE \bf
Control Barrier Functions: Theory and Applications
}

\author{Aaron D. Ames$^{1}$, Samuel Coogan$^{2}$, Magnus Egerstedt$^{3}$,\\ Gennaro Notomista$^{4}$, Koushil Sreenath$^{5}$, and Paulo Tabuada$^{6}$%
\thanks{$^{1}$ Mechanical and Civil Engineering and Control and Dynamical Systems, California Institute of Technology, Pasadena CA 91125, U.S.A, {\tt\small ames@caltech.edu}}%
\thanks{$^{2}$ Electrical and Computer Engineering and Civil and Environmental Engineering, Georgia Institute of Technology, Atlanta GA 30332, U.S.A, {\tt\small sam.coogan@gatech.edu}}%
\thanks{$^{3}$ Electrical and Computer Engineering, Georgia Institute of Technology, Atlanta GA 30332, U.S.A, {\tt\small magnus@gatech.edu }}%
\thanks{$^{4}$ Institute for Robotics \& Intelligent Machines, Georgia Institute of Technology, Atlanta GA 30332, U.S.A, {\tt\small g.notomista@gatech.edu}}%
\thanks{$^{5}$ Mechnical Engineering, Univ. of California, Berkeley CA 94720, U.S.A, {\tt\small koushils@berkeley.edu}}%
\thanks{$^{6}$ Electrical and Computer Engineering, UCLA, Los Angeles CA 90095, U.S.A, {\tt\small tabuada@ucla.edu}}%
}

\begin{document}

\maketitle
\thispagestyle{empty}
\pagestyle{empty}

\begin{abstract}
\subfile{sections/abstract}
\end{abstract}

\section{Introduction}
\label{sec:introduction}

\subfile{sections/1_introduction.tex}

\section{Foundations of Control Barrier Functions}
\label{sec:foundations}
\subfile{sections/2_foundations.tex}

\section{CBFs for Systems with Actuation constraints}
\label{sec:actuationconstraints}

\subfile{sections/4_actuationconstraints.tex}

\section{Exponential Control Barrier Functions}
\label{sec:exponential}
\subfile{sections/5_exponential.tex}

\section{Applications: CBFs for Robotic Systems}
\label{sec:robotics}
\subfile{sections/6_robotics.tex}

\subsection{Long Duration Autonomy}
\label{sec:autonomy}
\subfile{sections/7_autonomy.tex}

\section{Conclusions}
\subfile{sections/conclusions}

\balance

\bibliographystyle{IEEEtran}
\bibliography{bib/IEEEabrv,bib/ecc2019tutorialbibliography}

\end{document}

%% file: sections/abstract.tex
This paper provides an introduction and overview of recent work on control barrier functions and their use to verify and enforce safety properties in the context of (optimization based) safety-critical controllers. We survey the main technical results and discuss applications to several domains including robotic systems.

%% file: sections/1_introduction.tex
It is easy to agree that any engineered system should be designed to be \emph{safe}. In fact, the term \emph{safety-critical} system is many times used to distinguish those systems for which safety is a major design consideration. But what exactly is \emph{safety}? How do we define it and how can we design systems to achieve it? The notion of safety was first introduced in 1977 in the context of program correctness by Leslie Lamport~\cite{LL77} and formalized in~\cite{LL84}, see also~\cite{HS85}. Intuitively, safety requires that ``bad'' things do not happen while liveness requires that ``good'' things eventually happen, e.g., asymptotic stability can be seen as an example of a liveness property in the sense that an asymptotically stable equilibrium point is eventually reached. Dually, invariance can be seen as an example of a safety property in the sense that any trajectory starting inside an invariant set will never reach the complement of the set, describing the locus where bad things happen. Based on the identification of liveness with asymptotic stability and safety with invariance, it can be argued that safety has received much less attention in control theory than liveness. Moreover, the notion of Lyapunov function has played a predominant role in the investigation of liveness properties. 

The objective of this paper is to refocus the discussion on safety by introducing control barrier functions that play a role equivalent to Lyapunov functions in the study of liveness properties. There are two main reasons driving a surge in research related to safety and control barrier functions: 1) the recent interest in autonomous systems has brought safety to the forefront of systems' design. In particular, autonomous systems are expected to operate in unknown and unstructured environments which makes it considerably harder to enforce safety properties; 2) the recent introduction of control barrier functions suggests that many control design techniques based on Lyapunov and control Lyapunov functions can be suitably transposed to address safety considerations. Hence, we have both the societal need for safety as well as the tools to raise safety to the same level of maturity than liveness in the design of control systems.

\subsection{Brief History of Barrier Functions}

The study of safety in the context of dynamical systems dates back to the 1940's when Nagumo provided necessary and sufficient conditions for set invariance \cite{nagumo1942lage} (see \cite{blanchini1999set} for a
more detailed historical account, and \cite{abraham2012manifolds} for a modern proof).  In particular, given a dynamical system $\dot{x} = f(x)$ with $x \in \R^n$, assuming that the safe set $\C$ is the superlevel set of a smooth function $h : \R^n \to \R $, i.e., $\C = \{ x \in \R^n ~:~ h(x) \geq 0\}$, and that $\frac{\partial h}{\partial x}(x)\neq 0$ for all $x$ such that $h(x)=0$, then Nagumo's Theorem gives necessary and sufficient conditions for set invariance based upon the derivative of $h$ on the boundary of $\C$: 
$$
\C ~\mathrm{is}~ \mathrm{invariant} \quad \Leftrightarrow \quad \dot{h}(x) \geq 0 ~ \forall ~ x \in \partial \C.
$$
These conditions have been independently re-discovered on multiple occasions; in particular, around the 1970s by Bony and Brezis \cite{bony1969principe,brezis1970characterization} 
(the proof in \cite{abraham2012manifolds} follows Brezis). 

In the 2000's we saw another change of perspective brought by the need to verify hybrid systems. {\it Barrier certificates} were introduced as a convenient tool to formally prove safety of nonlinear and hybrid systems~\cite{prajna2004safety,prajna2006barrier}; these results, again, seemed to independently discover Nagumo's theorem. The choice of the term ``barrier'' was motivated by its use in the optimization literature where barrier functions are added to cost functions to avoid undesirable regions.  In the case of barrier certificates, one considers an unsafe set $\C_u$ and a set of initial conditions $\C_0$ together with a function $B : \R^n \to \R$ where $B(x) \leq 0$ for all $x \in \C_0$ and $B(x) > 0$ for all $x \in \C_u$.  Then $B$ is a barrier certificate if
$$
\dot{B}(x) \leq 0 \quad \Rightarrow \quad \C ~\mathrm{is}~ \mathrm{invariant} 
$$
In the notation for $\C$ above, by picking the safe set to be the complement of the unsafe set $\C = \C_u^c$, with $B(x) = -h(x)$ the barrier certificate conditions become: $\dot{h}(x) \geq 0$ which implies that $\C$ is invariant.  Therefore, these conditions reduce to those of Nagumo's theorem on the boundary.
Importantly, the necessity of barrier certificates were studied \cite{prajna2005necessity} along with their extension to a stochastic setting \cite{prajna2007framework}.

As a means to extend the safety guarantees beyond the boundary of the set, there have been a variety of approaches that can be best described as ``Lyapunov-like.''  That is, Lyapunov functions yield invariant level sets so, if these level sets are contained in the safe set one can guarantee safety---importantly, these conditions can be applied over the entire set and not just on the boundary.  In this case, as developed in \cite{tee2009barrier}, one constructs a ``barrier Lyapunov function" $B$ much as above but with the additional requirement that it is, for all intents and purposes, positive definite.  Then, by enforcing the condition that $\dot{B} \leq 0$ over the set $\C$, it ensures invariance of this set and thus safety.   The major limitation is that, while these conditions ensure safety they also enforce invariance of \emph{every} level set.  Thus, they are overly strong and conservative.

While the above results addressed closed dynamical systems, i.e., systems without inputs, the work on \emph{viability theory}~\cite{aubin2009viability,aubin1990survey,aubin2011viability} extended them to open dynamical systems, e.g., control systems given by $\dot{x} = f(x) + g(x) u$ for $u \in U \subset \R^m$. This required moving from invariant sets to controlled invariant sets: sets that can be made invariant by suitably designing a controller.

The notion of a barrier certificate was extended to a ``control'' version to yield the first definition of a ``control barrier function'' \cite{wieland2007constructive}---although this definition is different than the one considered in this paper.  In particular, given a control system and a safe set $\C$ as defined above by a function $h$, the conditions in \cite{wieland2007constructive} are effectively: 
$$
\exists ~ u ~\mathrm{s.t.} ~ ~ \dot{h}(x,u) \geq 0 \quad \Rightarrow \quad 
\C ~ \mathrm{is} ~ \mathrm{invariant}
$$
These ideas were built upon so as to explicitly combine barrier functions with control Lyapunov functions \cite{romdlony2014uniting}---this was done contemporaneously with the development of the methods presented in this paper which use optimization based controllers to unify Lyapunov and barrier functions.  
In particular, as further developed in \cite{romdlony2016stabilization}, conditions were given on creating ``control Lyapunov barrier functions'' that jointly guarantee safety and stability. 
Yet, in these cases the conditions in the end reduce to enforcing $\dot{h}(x,u) \geq 0$.  However, these conditions are stronger than necessary, and thus motivate the ``modern'' version of control barrier functions.   

The aforementioned methods all led to the most recent formulation of certificates of safety, termed {\it control barrier functions}, as recognition of the historical developments outlined above---these were first introduced in \cite{ames2014control}, and later refined in \cite{ames2017cbf}. In particular, the idea was to extend the barrier function conditions (e.g., those discovered by Nagumo) to the entirety of the safe set.  For a control system, and a safe set $\C$ defined by a function $h$, this new form of control barrier functions are defined by the condition: 
$$
\exists ~ u ~\mathrm{s.t.} ~ ~ \dot{h}(x,u) \geq - \alpha(h(x)) \quad \Leftrightarrow \quad 
\C ~ \mathrm{is} ~ \mathrm{invariant}
$$
for $\alpha$ an (extended) class $\classK$ function.  
Importantly, this condition is necessary and sufficient (for compact sets) and thus is minimally restrictive. Finally, because these conditions are true over the entire set $\C$ they give a way to synthesize safe controllers---in this case, through the use of optimization-based control methods that modify the desired controller again in a minimally invasive fashion. This formulation, therefore, provides a foundational framework for safety-critical control.  

The utility of this new formulation of control barrier functions is evidenced by the application domains it has been applied to since its inception, including: automotive systems \cite{xu2015robustness,Xu2017_ccta_lk_asr,Xu:2018tk}, mulit-robot systems \cite{borrmann2015control,wang2017safety,pickem2017robotarium}, quadrotors \cite{wu2016safety,wang2017safe} and robotic systems including walking robots \cite{CBF:Ames:CBFBiped:ACC15,CDC2016_3DWalking_SteppingStones,gurriet2018towards}, to name a few.  Additionally, it allows for the unification of safety (via a control barrier function) and stability (via a control Lyapunov function) in the context of an optimization based controller---in fact, it was optimization based controllers using control Lyapunov functions that motivated the development of this new form of barrier function. This formulation of control barrier functions will be the focus of this paper, as motivated by the conceptual connections with control Lyapunov functions together with a recognition of the basic differences between control barrier and Lyapunov functions.

\subsection{Overview of Paper}

Building upon the history of barrier functions, and motivated by the new developments, this paper aims to establish the basic theory of safety-critical control and highlight some important applications.

\newsec{Theory:}  We begin in Section \ref{sec:foundations} by establishing the foundations of control barrier functions.  This is motivated from the perspective of stabilization with control Lyapunov functions, leading to the ``dual'' of stability: safety as enforced by control barrier functions.  The properties of these functions are discussed, along with the synthesis of optimization-based controllers.  In Section \ref{sec:actuationconstraints}, the application of CBFs to systems with actuation constraints is considered.  Finally, in Section \ref{sec:exponential}, the extension of CBFs to constraints with higher relative degree is considered.  

\newsec{Application:}  The discussion of the application of CBFs begins in Section \ref{sec:robotics} with the consideration of robotic systems.  In particular, we begin by considering the ``stepping stone'' problem, wherein a robot must walk safely on a series of stepping stones.  This is followed by a brief discussion of the experimental implementation of barriers in the context of automotive safety systems and dynamic robotic systems.  Additionally, the application of CBFs in the context of long duration autonomy is formulated and demonstrated experimentally.

%% file: sections/2_foundations.tex
In this section, we introduce the fundamentals of control barrier functions.  That is, we introduce safety, safety sets, and a means in which to enforce safety in a minimally invasive fashion.  To motivate these considerations, we will begin by reviewing control Lyapunov functions (CLFs) and discuss how they can be used to synthesize controllers that enforce stability.  This naturally leads to the ``dual'' for safety: control barrier functions (CBFs).  We will formulate optimization based controllers from CBFs and conclude by describing how they can be unified with CLFs. 

Throughout this paper, we will suppose that we have a nonlinear affine control system:
\begin{align}
\label{eqn:controlsys}
\dot{x} = f(x) + g(x) u,
\end{align}
with $f$ and $g$ locally Lipschitz, $x \in D \subset \R^n$ and $u \in U \subset \R^m$ is the set of admissible inputs.

\subsection{Motivation: Control Lyapunov Functions}  
\label{sec:CLFs}
To motivate safety for systems of this form, and hence control barrier functions, we begin by considering the familiar objective of stabilizing the system. 
Suppose we have the control objective of (asymptotically) stabilizing the nonlinear control system \eqref{eqn:controlsys} to a point $x^* = 0$, i.e., driving $x(t) \to 0$.  In a nonlinear context, this can be achieved---and, in fact, understood---by equivalently finding a feedback control law that drives a positive definite function, $V : D  \subset \R^n \to \Rplus$, to zero.  That is, if
\begin{eqnarray}
\label{eqn:motivatingCLF}
\exists ~ u = k(x) \quad \mathrm{s.t.} \quad \dot{V}(x,k(x)) \leq - \gamma(V(x)),
\end{eqnarray} 
where 
$$
\dot{V}(x,k(x)) = L_f V(x) + L_g V(x) k(x),
$$
then the system is stabilizable to $V(x^*) = 0$, i.e., $x^* = 0$.  Note that here $\gamma: \Rplus \to \Rplus$ is a class $\classK$ function defined on the entire real line for simplicity, i.e., $\gamma$ maps zero to zero, $\gamma(0) = 0$, and it is strictly monotonic: for all $r_1,r_2 \in \Rplus$, $r_1 < r_2$ implies that $\gamma(r_1) < \gamma(r_2)$.  Thus, the process of stabilizing a nonlinear system can be understood as finding an input that creates a one-dimensional stable system given by the Lyapunov function: $\dot{V} \leq - \gamma(V)$, wherein the comparison lemma (see, e.\,g., \cite{khalil2015nonlinear}) implies that the full-order nonlinear system \eqref{eqn:controlsys} is thus stable under the control law $u = k(x)$.

The above observations motivate the notion of a {\it control Lyapunov function} wherein a function $V$ is shown to stabilize the system without the need to explicitly construct the feedback controller $u = k(x)$. That is, as first observed by Sontag and Artstein \cite{Sontag:firstCLF,Sontag:universal,artstein1983stabilization}, we only need a controller to exist that results in the desired inequality on $\dot{V}$.  Concretely, $V$ is a \underline{control Lyapunov function (CLF)} if it is positive definite and satisfies: 
\begin{align}
\label{eqn:CLFdef}
\inf_{u \in U}\left[ L_f V(x) + L_g V(x) u\right] \leq - \gamma(V(x)), 
\end{align}
where $\gamma$ is again a class $\classK$ function.  The importance of this definition is that it allows for us to consider the set of all stabilizing controllers for every point $x \in D$: 
\begin{align}
\label{eqn:Kclf}
\Kclf(x) := \{ u \in U ~ : ~ L_f V(x) + L_g V(x) u \leq  -\gamma(V(x))\}.
\end{align}
This is an affine constraint in $u$ and thus will allow for the formulation of optimization based controllers.  It also elucidates conditions on when $V$ is a CLF; for example, if $U = \R^m$, it is easy to verify that
\begin{eqnarray}
L_g V(x) = 0 \quad & \implies & \quad L_f V(x) \leq - \gamma(V(x)) \nonumber\\
& \implies & \quad \Kclf(x) \neq \emptyset \nonumber
\end{eqnarray}
and thus there are stabilizing controllers.  More generally, we have the following central stabilization result for CLFs \cite{ames2014rapidly}. 

\begin{theorem}
For the nonlinear control system \eqref{eqn:controlsys}, if there exists a control Lyapunov function $V : D \to \Rplus$, i.e., a positive definite function satisfying \eqref{eqn:CLFdef}, then any Lipschitz continuous feedback controller $u(x) \in \Kclf(x)$ asymptotically stabilizes the system to $x^* = 0$. 
\end{theorem}

\subsection{Control Barrier Functions}

Unlike stability which involves driving a system to a point (or a set), safety can be framed in the context of enforcing invariance of a set, i.e., not leaving a {\it safe set}.  In particular, we consider a set $\C$ defined as the {\it superlevel set} of a continuously differentiable function $h : D \subset \R^n \to \R$, yielding:
\begin{eqnarray}
\label{lect24:eqn:superlevelsetC}
\C &=& \{ x \in D \subset \R^n : h(x) \geq 0\}, \nonumber\\
\label{lect24:eqn:superlevelsetC2}
\partial \C &=& \{ x \in  D \subset  \R^n : h(x) = 0\}, \\
\label{lect24:eqn:superlevelsetC3}
\mathrm{Int}(\C) &=& \{ x \in  D \subset  \R^n : h(x) > 0\}. \nonumber
\end{eqnarray}
We refer to $\C$ as the \underline{safe set}.

\newsec{Safety.}  Let $u = k(x)$ be a feedback controller such that the resulting dynamical system
\begin{eqnarray}
\label{eqn:dynamicalsystem}
\dot{x} = f_{\mathrm{cl}(x)} := f(x) + g(x) k(x)
\end{eqnarray}
is locally Lipschitz. To formally define safety, due to the locally Lipschitz assumption, for any initial condition $x_0 \in D$ there exists a maximum interval of existence $I(x_0) = [0,\tau_{\mathrm{max}})$ such that $x(t)$ is the unique solution to \eqref{eqn:dynamicalsystem} on $I(x_0)$; in the case when $f_{\mathrm{cl}}$ is forward complete \cite{khalil2015nonlinear}, $\tau_{\mathrm{max}} = \infty$.  This allows us to define {\it safety}:

\begin{definition}
The set $\C$ is \underline{forward invariant} if for every $x_0 \in \C$, $x(t) \in \C$ for $x(0) = x_0$ and all $t \in I(x_0)$.  The system \eqref{eqn:dynamicalsystem} is \underline{safe} with respect to the set $\C$ if the set $\C$ is forward invariant.
\end{definition}

\newsec{Control Barrier Functions (CBFs).}  
Using control Lyapunov functions as motivation, we wish to generalize to the concept of safety.  Yet, one must be careful about directly generalizing Lyapunov (as done, in particular, in \cite{Tee}).  If there exists a CLF $V$ such that $V(x) = 0 \implies x \in \C$ and $V$ has a superlevel set $\Omega_c = \{ x \in D ~: ~ V(x) \leq c\} \subset \C$, then the corresponding controllers in \eqref{eqn:Kclf} will render $\Omega_c$ invariant, and hence $\C$ safe.  Nevertheless, this is overly restrictive as it would render every sublevel set invariant, i.e., $\Omega_{c'}$ for all $c' < c$.  Rather, we wish to enforce set invariance without requiring a positive definite function, i.e., for $h$ to be a control barrier function it should render $\C$ invariant but not its sublevel sets.  

This motivates the formulation of control barrier functions.   Before defining these, we note that an \underline{extended class $\classKinfty$ function} is a function $\alpha : \R \to \R$ that is strictly increasing and with $\alpha(0) = 0$; that is, extended class $\classKinfty$ functions are defined on the entire real line: $\R = (-\infty,\infty)$.
This allows us to define \cite{ames2017cbf,xu2015robustness}:

\begin{definition}
\label{def:csf}
Let $\C \subset D \subset \R^n$ be the superlevel set of a continuously differentiable function $h: D \to \R$, then $h$ is a \underline{control barrier function (CBF)} if there exists an extended class $\classKinfty$ function $\alpha$ such that for the control system \eqref{eqn:controlsys}:
\begin{align}
\label{eqn:cbf:definition}
\sup_{u \in U}  \left[ L_f h(x) + L_g h(x) u \right] \geq - \alpha(h(x)).
\end{align}
for all $ x \in D$.
\end{definition}

\begin{remark}
{\rm 
Note that, as discussed in Section \ref{sec:introduction}, the first notion of a control barrier function \cite{ames2014control} was defined in terms of what are now termed {\it reciprocal} barrier functions.  These blow-up on the boundary, hence the use of the term ``barrier'': 
\begin{eqnarray}
\inf_{x \in \mathrm{Int}(\mathcal{C})}  B(x)  \geq  0 , \qquad
\lim_{x \to \partial\mathcal{C}}  B(x)   =  \infty.   \label{eqn:rbarrierproperty}
\end{eqnarray}
wherein the control barrier function condition \eqref{eqn:cbf:definition} becomes: 
\begin{align}
 \label{eq:rBinf}
& \inf_{u \in U}  \left[ L_f B(x) + L_g B(x) u  \right] \leq  \alpha\left( \frac{1}{B(x)} \right).
\end{align}
This class of barrier functions can be more suitable for some applications, but typically barrier functions, $h$, are preferable since they are well defined outside of $\C$. 
}
\end{remark}

\begin{remark}
{\rm 
The idea of extending set invarience conditions, i.e., the condition that $\dot{h} \geq 0$ for all $x \in \partial \C$, to all of $\C$ was first considered in \cite{aubin2009viability} in the form of the following condition: $\dot{h} \geq - h$ for all $x \in \C$.  This can be viewed as a very special case of a CBF wherein $\alpha(r) = r$ in \eqref{eqn:cbf:definition}. 
}
\end{remark}

\newsec{Guaranteed Safety via CBFs.}  We can consider the set consisting of all control values that render $\C$ safe:
\begin{align}
\label{eqn:CBFpart1:safecontrollers}
\Kcbf(x) = \{ u \in U : L_f h(x) + L_g h(x) u + \alpha(h(x)) \geq 0\}.
\end{align}
That is, as in the case of CLFs, we can quantify the set of all control inputs at a point $x \in D$ that keep the system safe. 

The main result of \cite{ames2017cbf}, and the main result with regard to control barrier functions, is that the existence of a control barrier function implies that the control system is safe:

\begin{theorem}
\label{thm:cbf}
{\it Let $\mathcal{C} \subset \R^n$ be a set defined as the superlevel set of a continuously differentiable function $h: D \subset \R^n \to \R$.  If $h$ is a control barrier function on $D$ and $\frac{\partial h}{\partial x}(x)\neq 0$ for all $x\in\partial \C$, then any Lipschitz continuous controller $u(x) \in \Kcbf(x)$ for the system \eqref{eqn:controlsys} renders the set $\C$ safe.  Additionally, the set $\C$ is asymptotically stable in $D$. }
\end{theorem}

\begin{remark}
{\rm 
The condition that the gradient of $h$ not vanish on the boundary is equivalent to requiring that $0$ is a regular value of $h$ \cite{abraham2012manifolds}.  Note that this condition was not explicitly stated in \cite{ames2017cbf}, but the proof of this result utilizes Nagumo's theorem \cite{nagumo1942lage} which requires this regularity condition \cite{abraham2012manifolds}.  
}
\end{remark}

\begin{remark}
{\rm 
It is important to stress that this result not only guarantees that the safe set $\C$ is invariant, but makes the set $\C$ asymptotically stable.  This has beneficial consequences with regard to practical implementation.  While a system will not formally leave the safe set $\C$, noise and modeling errors might force the system to leave this set.  As a result of the main CBF theorem, controllers in $\Kcbf(x)$ will drive the system back to the set $\C$. 
}
\end{remark}

\newsec{Necessity for Safety.}  Finally, we note that control barrier functions provide the strongest possible conditions for safety in that they are necessary and sufficient given reasonable assumptions on $\C$ \cite{ames2017cbf}:

\begin{theorem}
Let $\C$ be a compact set that is the superlevel set of a continuously differentiable function $h : D \to \R$ with the property that $\frac{\partial h}{\partial x}(x)\neq 0$ for all $x\in \partial \C$.  If there exists a control law $u = k(x)$ that renders $\C$ safe, i.e., $\C$ is forward invariant with respect to \eqref{eqn:dynamicalsystem}, then $h|_{\C} : \C \to \R$ is a control barrier function on $\C$. 
\end{theorem}

\subsection{Optimization Based Control}
\label{subsec:optbasedcontrol}

Having established that control barrier functions give (necessary and sufficient) conditions on safety, the question becomes: how does one synthesize controllers?  Importantly, we wish to do so in a minimally invasive fashion, i.e., modify an existing controller in a minimal way so as to guarantee safety.  This naturally leads to optimization based controllers: 

\newsec{Safety-Critical Control.}  Suppose we are given a feedback controller $u = k(x)$ for the control system \eqref{eqn:controlsys} and we wish to guarantee safety.  Yet it may be the case that $k(x) \notin \Kcbf(x)$ for some $x \in D$.  To modify this controller in a minimal way so as to guarentee safety, we start by noticing that the conditions on safety given in \eqref{eqn:CBFpart1:safecontrollers} are affine in $u$.  Thus, we can consider the following Quadratic Program (QP) based controller that finds the minimum perturbation on $u$:   
\begin{align}
\label{lectCBFpart1:eqn:mopt}
\tag{CBF-QP}
u(x) = \underset{u \in \R^{m}}{\operatorname{argmin}} &  \quad \frac{1}{2} \| u - k(x) \|^2  \\
\mathrm{s.t.} &  \quad L_f h(x) + L_g h(x) u \geq - \alpha(h(x))  \nonumber
\end{align}
where here we assumed that $U = \R^m$.  Thus, when there are no input constraints, since we have a single inequality constraint the CBF-QP has a closed-form solution (per the KKT conditions \cite{boyd2004convex}) given by the {\it min-norm controller}; this was first utilized in the context of CLFs \cite{freeman2008robust,ames2014rapidly}.  

\newsec{Unifying with Lyapunov.}  The QP based formulation of safety-critical controllers suggests a means in which to unify safety and stability.  In fact, optimization-based controllers were first utilized in the context of CLFs exactly for the purpose of multi-objective nonlinear control \cite{ames2013towards}, e.g., combining stability with torque constraints \cite{galloway2015torque}.  Concretely, we consider the following QP based controller: 
\begin{align}
\label{lect24:eqn:QPCLFCBF}
\tag{CLF-CBF QP}
u(x) =    
\underset{(u , \delta ) \in \R^{m+1}}{\operatorname{argmin}} &  \quad \frac{1}{2}
u^T H(x)u + p \delta^2 \\
%\label{lect24:eqn:CLFcon}
%\tag{CLF}
\mathrm{s.t.} &  \quad L_f V(x) + L_g V(x) u \leq - \gamma(V(x)) + \delta \nonumber\\
%\label{lect24:eqn:CBFcon}
%\tag{CBF}
 &  \quad L_f h(x) + L_g h(x) u \geq  -\alpha(h(x)) \nonumber
\end{align}
where here $H(x)$ is any positive definite matrix (pointwise in $x$), and $\delta$ is a relaxation variable that ensures solvability of the QP as penalized by $p > 0$ (i.e., to ensure the QP has a solution one must relax the condition on stability to guarantee safety). In \cite{ames2017cbf} it was established that this controller is Lipschitz continuous.

%% file: sections/4_actuationconstraints.tex
Consider again the nonlinear affine control system \eqref{eqn:controlsys} and assume there exists an \emph{allowable set} of states $A=\{x\in D:\rho(x)\geq 0\}$ defined via some \emph{performance function} $\rho:{D}\to \mathbb{R}$. Our objective is to construct a CBF $h:{D}\to \mathbb{R}$ such that
\begin{equation}
  \label{eq:safe_allow}
  \{x\in {D}:h(x)\geq 0\}\subseteq \{x\in D:\rho(x)\geq 0\},
\end{equation}
that is, such that the safe set $\mathcal{C}$, corresponding to the superlevel set of the CBF $h$, is contained within the set of allowed states $A$. Of course, it may be possible to take $h(x)=\rho(x)$ if this choice satisfies \eqref{eqn:cbf:definition} for an appropriate function $\alpha$, in which case our objective is met. 

However, in this section, we focus on the case when ${A}$ cannot be rendered invariant and instead we must find a safe subset that is a strict subset of the allowable set. The inability of ${A}$ itself to be rendered forward invariant could be due to, \emph{e.g.}, a control set ${U}$ that restricts the available control actions or due to dynamics with higher relative degree; an alternative approach to accommodate the latter is proposed in Section \ref{sec:exponential}.

We assume that a locally Lipschitz \emph{nominal controller} $\beta:D\to U$ (called \emph{nominal evading maneuver} in \cite{Squires:2018nr}) is known. Intuitively, $\beta$ encapsulates a controller that, for some initial conditions, is expected to keep the system within the allowable set, although no guarantees on the ability of $\beta$ to ensure safety are required \emph{a priori}. For example, for an autonomous mobile agent, $\beta$ might be a swerving maneuver or a rapid deceleration maneuver. 

For any $t\geq 0$ and $x\in D$, let $\phi_\beta(t,x)$ denote the state of the control system \eqref{eqn:controlsys} at time $t$ when $\beta$ is used as input and the system is initialized at $x$, that is, $\phi_\beta(t,x)$ satisfies $\dot{\phi}_\beta(t,x)=f(\phi_\beta(t,x))+g(\phi_\beta(t,x))\beta(\phi_\beta(t,x))$ with initial condition $\phi_\beta(0,x)=x$.

A barrier function can be computed from $\rho$ and $\beta$ as
\begin{align}
  \label{eq:h_allow}
  h(x)=\inf_{\tau\in[0,\infty)}\rho(\phi_\beta(\tau,x)),
\end{align}
that is, the barrier $h$ is constructed by assigning to each point $x\in D$ the infimum value of the performance function $\rho$ attained along the trajectory initialized at $x$ when the nominal control strategy $\beta$ is used. Under mild conditions on $\rho$ and $\beta$, $h$ is indeed a CBF \cite{Squires:2018nr}.

\begin{theorem}
Let $\rho(x)$ be a continuously differentiable performance function and let $\beta(x)$ be a nominal controller such that $f(x)+g(x)\beta(x)$ is continuously differentiable. Define $h$ as in \eqref{eq:h_allow} with $\mathcal{C}$ the corresponding superlevel set of $h$ and suppose for each $x$ there exists a unique $x^*$ such that $h(x)=\rho(x^*)$ and $\phi_\beta(\tau,x)=x^*$ for some $\tau\geq 0$. Then
\begin{enumerate}
\item $h$ is a CBF;
\item $\mathcal{C} \subseteq A$, that is, the safe set is a subset of the allowable set; and
\item $\beta(x)\in \Kcbf(x)$ for all $x\in \mathcal{C}$.
\end{enumerate}
\end{theorem}

In some cases, computing $h$ given in \eqref{eq:h_allow} is possible in closed form; see \cite{Squires:2018nr} for examples. 

Alternatively, one could approximate $h$ by simulating the system trajectory for a finite horizon and computing the infimum in \eqref{eq:h_allow} numerically. However, notice that to use $h$ in a resulting quadratic program as in \eqref{lectCBFpart1:eqn:mopt} requires computing the gradient of $h$, thus such an approach would also require numerically approximating the gradient of $h$, and therefore this approach  becomes computationally challenging as the dimension of the system grows.

Another approach is to parameterize $h$ and search for a potentially conservative CBF satisfying \eqref{eq:safe_allow}. For example, we could parameterize $h$ as a fixed degree polynomial and use {sums-of-squares (SOS) programming} \cite{Parrilo:2001uq} to enforce the required conditions on $h$. To this end, a polynomial $s(x)$ is a \underline{SOS polynomial} if $s(x)=\sum_{i=1}^r(g_i(x))^2$ for some polynomials $g_i(x)$ for $i=1,\ldots,r$. Let $\Sigma[x]$ denote the set of SOS polynomials in $x$. The following Proposition is closely related to results presented in \cite{Xu:2018tk, wang2018permissive}.

\begin{proposition}
\label{prop:SOS}
  Given the affine control system \eqref{eqn:controlsys}, assume $f(x)$ and $g(x)$ are polynomials. Let $\rho(x)$ be a polynomial performance function and let $\beta(x)$ be a polynomial nominal controller. A polynomial $h(x)$ is a CBF if there exists positive constants $a>0, \epsilon>0$ and SOS polynomials $s_1(x)$, $s_2(x)$ such that
  \begin{align}
    \label{eq:3}
-h(x)-\epsilon+s_1(x)\rho(x)&\in\Sigma[x],\\
\label{eq:3-2} L_fh(x)+L_gh(x)\beta(x)+a h(x)-s_2(x) h(x)&\in\Sigma[x].
  \end{align}
Moreover, $\mathcal{C} \subseteq A$ and $\beta(x)\in \Kcbf(x)$ for all $x\in \mathcal{C}$.
\end{proposition}

Condition \eqref{eq:3} is sufficient for ensuring that $h(x)< 0$ for all $x$ such that $\rho(x)< 0$, thereby implying $\mathcal{C}\subseteq A$. Likewise, \eqref{eq:3-2} is sufficient for ensuring that $L_fh(x)+L_gh(x)\beta(x)+a h(x)\geq 0$ for all $x\in\mathcal{C}$. Since $\beta(x)\in U$ for all $x\in D$, this in turn implies \eqref{eqn:cbf:definition} with the choice $\alpha(s)=a s$.

There exist efficient computational toolboxes that convert certain SOS constraints into semidefinite programs such as \cite{sostools}.  However, viewing  $h(x)$, $a$, $s_1(x)$, and $s_2(x)$ as decision variables in the above, the products $a h(x)$ and $s_2(x) h(x)$ are bilinear in the decision variables and prevent such a conversion. 

Nonetheless, a common approach for accommodating such bilinearities is to propose an iteration of constraints so that in each iteration, one element of each problematic product is fixed, i.\,e., in each iteration, either $a$ and $s_2(x)$ are fixed or $h(x)$ is fixed, leading to an efficient numerical procedure for finding a CBF $h$. For example, in \cite{Xu:2018tk}, a sequence of SOS programs is proposed to compute a CBF for lane-keeping and adaptive cruise control in an autonomous vehicle, and in \cite{wang2018permissive}, a sequence of SOS programs is proposed to compute a region of safe stabilization. 

Variants of the SOS-based approach proposed in Proposition \ref{prop:SOS} are possible and have been explored in related contexts, e.\,g., \cite{Xu:2018tk, wang2018permissive}. For example, it is possible to compute a new nominal controller after computing a barrier $h(x)$. Further, the constraints \eqref{eq:3}--\eqref{eq:3-2} can be augmented with an objective function that, e.\,g., seeks to maximize the volume of the safe set $\mathcal{C}$. In addition, it is possible to consider an allowable set characterized as the intersection of the superlevel sets of multiple performance functions by including a constraint like \eqref{eq:3} for each performance function.

%% file: sections/5_exponential.tex
In the previous sections we have seen how control barrier functions (CBFs) can be (i) used to enforce safety-critical constraints for nonlinear (control affine) systems, (ii) combined with control Lyapunov functions to arbitrate between stability and safety, and (iii) used for systems with actuator constraints.  While CBFs offer a powerful methodology, there is one critical restriction: the safety-critical constraints have been so far assumed to be of relative-degree one, i.e., the first time-derivative of the CBF has to depend on the control input.  However, this is a restrictive assumption that is typically not held for most safety constraints for robotic systems.  We therefore need a way to enforce arbitrarily high relative-degree safety constraints.  In this section, we introduce a special type of CBFs called Exponential CBFs that enable this functionality.

Control barrier functions for high-relative degree safety constraints were initially studied simultaneously in \cite{ACC2015_Geometric_CBFCLF, CBF:Ames:CBFBiped:ACC15}.  However, the results in \cite{ACC2015_Geometric_CBFCLF} only extended to position based safety constraints with relative-degree 2.  On the other hand, the results in \cite{CBF:Ames:CBFBiped:ACC15} extended to arbitrary high relative-degree using a backstepping based method.  However, backstepping based CBF design for higher relative-degree systems (greater than 2) is challenging and has not been attempted.  Building off the work in \cite{ACC2015_Geometric_CBFCLF}, exponential control barrier functions were first introduced in \cite{ACC2016_Exponential_CBF} as a way to easily enforce high relative-degree safety constraints.  The rest of this section provides an introduction to exponential CBFs.

\subsection{High Relative-Degree Safety Constraints}
Consider the nonlinear dynamical system in \eqref{eqn:controlsys} with initial condition $x_0$ with the goal to enforce the forward invariance of the safe set $\C$ defined in \eqref{lect24:eqn:superlevelsetC}.  However, unlike in earlier sections, we relax the relative-degree 1 assumption on $h(x)$ and assume $h(x)$ has arbitrarily high relative-degree $r \ge 1$.  This translates to the $r^\text{th}$ time-derivative of $h(x)$ being,
\begin{equation} \label{eq:ECBF_time-derivatve}
	h^{(r)}(x,u) = L_f^r h(x) + L_g L_f^{r-1} h(x) u,
\end{equation}
with $L_g L_f^{r-1} h(x) \ne 0$ and $L_g L_f h(x) = L_g L_f^2 h(x) = \cdots = L_g L_f^{r-2} h(x) = 0, \forall x \in D$.
Next, we define, 
\begin{equation}
	\eta_b(x) := \begin{bmatrix}
		h(x) \\ \dot h(x) \\ \ddot h(x) \\ \vdots \\ h^{(r-1)}(x)	
	\end{bmatrix} = 
	\begin{bmatrix}
		h(x) \\ L_f h(x) \\ L_f^2 h(x) \\ \vdots \\ L_f^{r-1} h(x)
	\end{bmatrix},
\end{equation} and assume for a given $\mu \in U_\mu \subset \mathbb{R}$, $u$ can be chosen such that $L_f^r h(x) + L_g L_f^{r-1} h(x) u = \mu$.  This choice of $u$ is possible since by the relative degree of $h(x)$ we have $L_g L_f^{r-1} h(x) \ne 0, \forall x$ and moreover $\mu$ is a scalar (while $u \in U \subset \mathbb{R}^m$).  With this, the above dynamics of $h(x)$ can be written as the linear system,
\begin{eqnarray} \label{eq:ECBF_linear_sys}
	\dot \eta_b(x) &=& F \eta_b(x) + G \mu, \nonumber \\
	h(x) &=& C \eta_b(x), 
\end{eqnarray}
where
\begin{eqnarray}
	F &=& \begin{bmatrix}
		0 & 1 & 0 & \cdots & 0 \\
		0 & 0 & 1 & \cdots & 0 \\
		\vdots & \vdots & \vdots & \ddots & \vdots \\
		0 & 0 & 0 & \cdots & 1 \\
		0 & 0 & 0 & \cdots & 0 \\
	\end{bmatrix}, \quad
	G = \begin{bmatrix}
		0 \\
		0 \\
		\vdots \\
		0 \\
		1
	\end{bmatrix}, \\
	C &=& \begin{bmatrix} 1 & 0 & \cdots & 0 \end{bmatrix}. \nonumber
\end{eqnarray}
Clearly, if we choose a state feedback style $\mu = -K_\alpha \eta_b(x)$, then $h(x(t)) = C e^{(F-GK_\alpha)t} \eta_b(x_0)$.  Moreover, by the comparison lemma, if $\mu \ge -K_\alpha \eta_b(x)$, then $h(x(t)) \ge C e^{(F-GK_\alpha)t} \eta_b(x_0)$.

We now have everything setup to define exponential control barrier functions.

\begin{definition} \label{def:ecbf}
	Given a set $\C \subset D \subset \mathbb{R}^n$ defined as the superlevel set of a $r$-times continuously differentiable function $h : D \to \mathbb{R}$, then $h$ is an \underline{exponential control barrier function (ECBF)} if there exists a row vector $K_\alpha \in \mathbb{R}^r$ such that for the control system \eqref{eqn:controlsys},
\begin{align}
	\label{eq:ecbf_definition}
	\sup_{u \in U}  \left[ L_f^r h(x) + L_g L_f^{r-1} h(x) u \right] \geq - K_\alpha \eta_b(x)
\end{align}
$\forall$ $ x \in \mathrm{Int}(\mathcal{C})$ results in $h(x(t)) \ge C e^{(F-GK_\alpha)t} \eta_b(x_0) \ge 0$ whenever $h(x_0) \ge 0$.
\end{definition}

\begin{remark}
	{\rm 
	Note that $K_\alpha$ in the above definition needs to satisfy certain specific properties.  As we will see, we will require $K_\alpha$ to make the closed-loop system matrix stronger than Hurwitz (total negative) and additionally satisfy a condition based on the initial conditions $\eta_b(x_0)$.  These will be presented in more detail in the subsequent subsection on designing ECBFs.
	}
\end{remark}

\begin{remark}
    {\rm 
	Note that when the relative-degree $r=1$, $-K_\alpha \eta_b(x)$ in \eqref{eq:ecbf_definition} reduces to $-\alpha h(x)$ with $\alpha > 0$.  Thus, Definition \ref{def:csf} defines a relative-degree 1 exponential CBF when $\alpha(h(x)) = \alpha h(x)$ (with a small abuse of notation), $\alpha > 0$.  In this sense, the above definition is a generalization of the definition of CBFs for higher relative-degree functions $h(x)$.
  }
\end{remark}

Given an ECBF, we can implement a controller that enforces the condition given in Definition~\ref {def:ecbf} by extending the optimization based control methodology presented earlier.  Concretely, we can consider the following QP based controller:
\begin{align}
	\label{eq:CLF_ECBF_QP}
	\tag{\small CLF-ECBF QP}
		u(x) =  & \underset{(u , \mu, \delta ) \in \R^{m+2}}{\operatorname{argmin}} 
			\quad \frac{1}{2} u^T H(x)u + p \delta^2 \\
		& \qquad \mathrm{s.t.} 
			 \quad   L_f V(x) + L_g V(x) u \leq - \gamma(V(x)) + \delta \nonumber\\
			& \hspace{1.18cm} \quad L_f^r h(x) + L_g L_f^{r-1} h(x) u = \mu \nonumber \\
			& \hspace{1.18cm} \quad \mu \geq -K_\alpha \eta_b(x). \nonumber
\end{align}

\subsection{Designing Exponential Control Barrier Functions}
\label{sec:design_ECBF}
In order to design an exponential CBF, we begin by noting that \eqref{eq:ECBF_linear_sys} is in controllable canonical form and if $K_\alpha = \begin{bmatrix} \alpha_1 & \cdots & \alpha_r \end{bmatrix}$ then the characteristic polynomial of $F-GK_\alpha$ is $\lambda^r + \alpha_r \lambda^{r-1} + \cdots + \alpha_2 \lambda + \alpha_1 = 0$, whose roots we will denote by $p_1, \cdots, p_r$.  Note that there is a well established relation between the coefficients of a polynomial and its roots.

We next define a family of functions $\nu_i : D \to \mathbb{R}$ and corresponding superlevel sets $\C_i$ for $i=0, \cdots, r,$ as follows:
\begin{align}
	\nu_0(x) &= h(x), & \C_0 &= \{x : \nu_0(x) \ge 0\}, \nonumber \\
	\nu_1(x) &= \dot \nu_0(x) + p_1 \nu_0(x), & \C_1 &= \{x : \nu_1(x) \ge 0\}, \nonumber \\
	\vdots & & \vdots \nonumber \\
	\nu_r(x) &= \dot \nu_{r-1}(x) + p_r \nu_{r-1}(x), & \C_r &= \{x : \nu_r(x) \ge 0\}. \nonumber
\end{align}
Note that $\C_0$ is identical to $\C$.  Our goal is to design $K_\alpha$ to ensure $\C$ is forward invariant.  We begin with the following result.

\begin{proposition}[\cite{ACC2016_Exponential_CBF}]
	For a given $i \in \{1,\cdots,r\}$, if $\C_i$ is forward-invariant then $\C_{i-1}$ is forward-invariant whenever $p_i > 0$ and $x_0 \in \C_i \cap \C_{i-1}$.
\end{proposition}

The above result follows from noting that under the given conditions when $x(t)$ reaches the boundary of $\C_{i-1}$, we have $\dot \nu_{i-1} \ge 0$ resulting in forward invariance of $C_{i-1}$.  The recursive application of the above proposition then motivates the following result:

\begin{theorem}[\cite{ACC2016_Exponential_CBF}]
	If $\C_r$ is forward-invariant and $x_0 \in \bigcap_{i=0}^r C_i$ then $\C$ is forward-invariant.
\end{theorem}

From the above results, for invariance of $\C$, we require two conditions for each $i$: (a) $p_i > 0$ and (b) $x_0 \in \C_i$.  The first condition on $p_i$ implies that the poles of the closed-loop $F-GK_\alpha$ need to be real and negative.  The second condition on $x_0$ and the definition of $\C_i$ implies we require $\nu_i(x_0) \ge 0 \iff \dot \nu_{i-1}(x_0) + p_i \nu_{i-1}(x_0) \ge 0 \iff p_i \ge - \frac{\dot \nu_{i-1}(x_0)}{\nu_{i-1}(x_0)}$.  Both these conditions can be achieved by choosing $K_\alpha$ as specified in the main result below.

\begin{theorem}[\cite{ACC2016_Exponential_CBF}]
	Suppose $K_\alpha$ is chosen such that $F-GK_\alpha$ is Hurwitz and total negative (resulting in negative real poles) and the eigenvalues satisfy $\lambda_i(F-GK_\alpha) \ge - \frac{\dot \nu_{i-1}(x_0)}{\nu_{i-1}(x_0)}$, then $\mu \ge -K_\alpha \eta_b(x)$ guarantees $h(x)$ is an exponential CBF. 
\end{theorem}

Thus, an exponential CBF can be designed using classical pole placement strategies from linear feedback theory.  The location of the poles is specified to be both real and negative as well as dependent on the higher time-derivatives of the barrier function at initial time.

%% file: sections/6_robotics.tex
Having seen the theoretical development of control barrier functions in the earlier sections, we will now present practical uses of CBFs in various robotic application domains.  Sections~\ref{sec:dynamicwalking_steppingstones} to \ref{subsec:segway} will introduce CBFs for single-agent robotic systems: we will look at three sufficiently different types of robotic systems, i.\,e. walking robots, cars, and Segways. Section~\ref{sec:autonomy} will introduce CBFs for multi-agent robotic systems.

\subsection{Dynamic Walking on Stepping Stones}
\label{sec:dynamicwalking_steppingstones}
\begin{figure}
	\centering
	\includegraphics[width=0.8\linewidth]{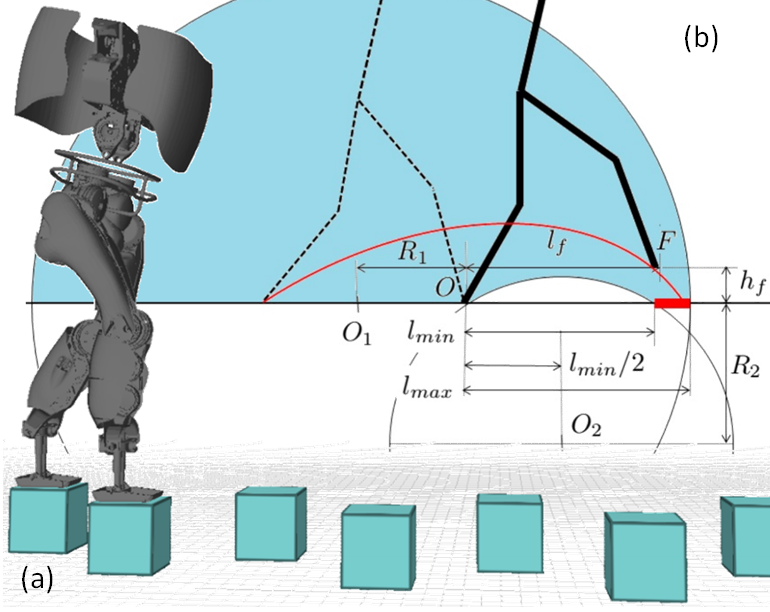}
\caption{(a) Foreground: The problem of dynamically walking over a terrain of stepping stones in 3D---a safety-critical problem.  (b) Background: Geometric depiction of step length foot placement constraint.  Here, $(O_1,R_1), (O_2, R_2)$ are the centers and radii of the outer and inner circles respectively, while $O$ is the position of the stance foot, $l_f, h_f$ denote the horizontal and vertical position of the swing foot with respect to the stance foot, and the red thick line between the distances of $l_{min}$ and $l_{max}$ from $O$ denote the stepping stone.}
	\label{fig:steppingstone}
\end{figure}

Legged robots are unique in the sense that these systems are able to locomote over discrete terrains - such as a terrain with steeping stones with discrete gaps between the steps (see Fig.~\ref{fig:steppingstone}a).  Precisely stepping on the footholds is critical and missing the foothold even by a few centimeters will cause a dramatic fall of the robotic system.  In this sense, stepping stones are examples of safety-critical control that have to be strictly enforced.  While this is challenging, in the preceding sections we have developed the theory to specifically attack such safety-critical problems.  Dynamic walking over stepping stones using CBFs was first demonstrated in \cite{ADHS2015_FootstepCBF}.  Here, we present results on the DURUS bipedal robot reported in \cite{CDC2016_3DWalking_SteppingStones}.

Legged systems are modeled as multi-domain hybrid systems with walking consisting of a single-support phase when one (stance) foot is in contact with the ground and an instantaneous double-support phase when the swing foot impacts ground. The single-support phase is modeled as a continuous-time differential equation while the double-support phase is modeled as an instantaneous impact due to the swing foot impacting on the ground.  The impact causes an instantaneous jump in the system state.  Mathematically, this is represented as the hybrid system
\begin{equation}
    \Sigma : \begin{cases}
        \dot x = f(x) + g(x) u, \quad x \notin S, \\
        x^+ = \Delta(x^-), \quad x \in S,
    \end{cases}
\end{equation}
with $S$ representing the switching surface that denotes swing foot contact with the ground.

For the above system, a hybrid zero dynamics (HZD) based approach (see \cite{WGCCM07} for details) is used to design a stable periodic orbit---representing walking---by means of an offline nonlinear constrained optimization, in order to find a set of outputs $y : \mathbb{R}^n \to \mathbb{R}^m$ that are then regulated by constructing a Lyapunov function $V(x) = \begin{bmatrix}y & \dot y\end{bmatrix}P\begin{bmatrix}y & \dot y\end{bmatrix}^T$ such that driving $V(x) \to 0$ results in driving the outputs to zero, resulting in stable walking.  This is achieved by the CLF based approach detailed in Section \ref{sec:CLFs}, with the difference for a hybrid system being that rapid exponential stability is sought through a RES-CLF \cite{ames2014rapidly} s.t. $\dot V(x, u) \le -\frac{1}{\epsilon} \gamma(V(x))$, where $0 < \epsilon < 1$.  This ensures that the controller contracts faster than the potential expansion that happens at impacts.  See \cite{ames2014rapidly,galloway2015torque} for more details.

\begin{figure}
	\centering
	\subfloat[][]{%
	    \includegraphics[width=0.8\linewidth]{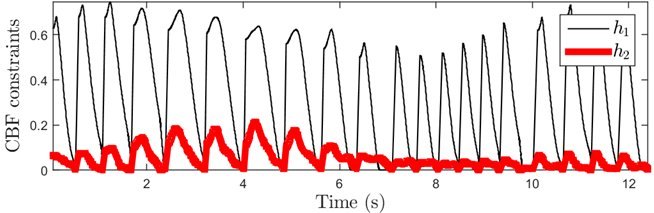}%
	    \label{fig:steplength_h1h2}%
    }\\
	\subfloat[][]{%
	    \includegraphics[width=0.8\linewidth]{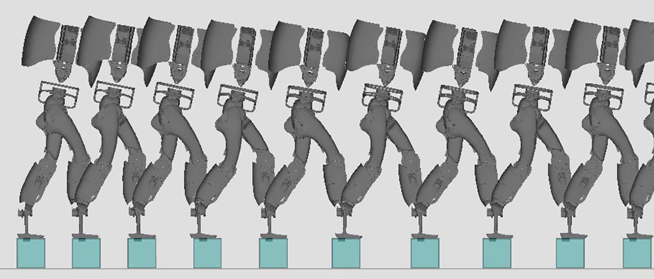}%
	    \label{fig:steplength_sim}%
    }
	\caption{Simulation results of dynamic walking over a terrain of stepping stones with varying step lengths. (a) Plots of the ECBFs $h_1, h_2$ being enforced. (b) Snapshots of walking from simulation. Simulation video: \url{https://youtu.be/yUSTraDn9-U}.}
	\label{fig:steppingstones_sim}
\end{figure}

Now, let us look into the problem of how we can guarantee the safety-critical constraint of precisely placing the feet on the stepping stone on each step.  In Fig.~\ref{fig:steppingstone}b, the start of the step is shown as the dotted stick-figure with the stance foot at $O$.  The goal is to move the swing leg and precisely impact the ground within the solid red foothold at the end of the step.  This is a constraint at the step end-time which can not be directly enforced as a barrier.  We convert this end-time constraint into a barrier constraint that is enforced point-wise in time.  In particular, if the swing foot position, denoted by $F$ in the Fig.~\ref{fig:steppingstone}b, is maintained within the outer circle (with center $O_1$ and radius $R_1$) and outside the inner circle (with center $O_2$ and radius $R_2$), then the foot follows the red trajectory and impacts the foothold at the end of step.  This can be formulated through enforcing the nonnegativity of the following CBFs:
\begin{eqnarray}
    h_1(x) &=& R_1 - O_1F(x) \ge 0, \nonumber \\
    h_2(x) &=& O_2F(x) - R_2 \ge 0, \nonumber
\end{eqnarray}
where $O_1F(x)$ and $O_2F(x)$ are the distances between the swing foot $F$ and the centers of the two circles at $O_1$ and $O_2$ respectively.
Since $h_i(x), i \in \{1,2\}$ are position constraints, they have relative-degree 2.  We thus use the tools of the exponential CBF to design $\alpha_{i,1}, \alpha_{i,2}$ and pick $u$ s.t., $L_f^2 h_i(x,u) + L_g L_f h_i(x) u \ge -\alpha_{i,1}h_i(x) - \alpha_{i,2}\dot h_i(x)$.  This results in enforcing $h_i(x) \ge 0$ resulting in dynamic walking on stepping stones.  Fig.~\ref{fig:steplength_h1h2} shows $h_1,h_2$ plotted against time to illustrate that they are non-negative.  Fig.~\ref{fig:steplength_sim} illustrates snapshots from simulation of walking over a stepping stone terrain with different step lengths.  This method can also be used to walk over a terrain of stepping stones with changing step width or step height.

\subsection{Automotive Systems: Automatic Cruise Control and Lane Keeping}
Our next example is from the automotive domain.  Many modern Advanced Driver Assistance Systems (ADAS) provide prime examples of safety-critical constraints. For instance, in Adaptive Cruise Control (ACC) the vehicle's speed is regulated to a user-set speed when there is no vehicle immediately ahead in the lane, yet if a vehicle is detected ahead then a safe following distance is maintained.  On the other hand, in Lane Keeping (LK) the vehicle's steering is controlled so as to maintain the vehicle within a lane.  Furthermore, two or more ADAS control modules can be simultaneously activated and designing provably correct controllers for simultaneous operation becomes critical; this subsection follows from \cite{Xu2017_ccta_lk_asr}, but see also \cite{ames2017cbf}.

In order to demonstrate adaptive cruise control and lane keeping in an experimental setting, we will consider a Khepera robot modeled as a unicycle model
\begin{equation}
    \begin{bmatrix}
        \dot p_x \\
        \dot p_y \\
        \dot v \\
        \dot \psi \\
        \dot \omega
    \end{bmatrix} =
    \begin{bmatrix}
        v \cos(\psi) - a \omega \sin(\psi) \\
        v \sin(\psi) + a \omega \cos(\psi) \\
        \frac{u_l}{m} - a\omega^2 \\
        \omega \\
        \frac{u_a}{I_z}
    \end{bmatrix},
\end{equation}
where $(p_x,p_y), \psi, v, \omega$ represent the 2D position, orientation, and longitudinal and angular velocities of the robot respectively, with $x \in \R^5$ the resulting state vector.  Further, $u_l$ is the longitudinal force and $u_a$ is the angular torque and serve as control inputs.  The mass and inertia are $m, I_z$ respectively and $a$ represents the distance from the center of the wheel-base to the point of interest $(p_x,p_y)$.  This model can be written as a nonlinear control affine system as given in \eqref{eqn:controlsys}.

As mentioned, adaptive speed regulation comprises of following a user-set speed when there is no vehicle ahead in the lane.  This will be formulated as a soft constraint through a CLF.  However,  when there is a vehicle ahead, the speed needs to be adaptively reduced so as to maintain a fixed time-headway based follow distance.  This will be enforced as a safety-critical constraint through the following CBF:
$$
h_{asr}(x) = D - \tau v_f.
$$
Here, $D$ is the distance to the vehicle ahead, $\tau$ is minimum time-headway to be maintained, and $v_f$ is the velocity of the vehicle (follower)---see \cite{ames2014control} for the derivation.

Similarly, the objective of lane keeping is to maintain the vehicle within the lane.  We need to enforce a safety-critical constraint of the form $y_{lat} \le d_{max}$, where $y_{lat}$ is the lateral distance w.r.t. the center of the lane and $d_{max}$ is the distance from the center of the lane to either end of the lane that captures the lane width. We enforce this safety constraint through the following CBF:
$$
h_{lk}(x) = d_{max} - \mathrm{sign}(v_{lat})y_{lat} -\frac{1}{2} \frac{v_{lat}^2}{a_{max}}. 
$$
Here, $a_{max}$ is the maximum lateral acceleration and $v_{lat}$ is the lateral velocity of the vehicle.  More details about the properties of this CBF are detailed in \cite{ames2017cbf,Xu2017_ccta_lk_asr}.

\begin{figure}
	\centering
	\subfloat[][]{%
	    \includegraphics[width=0.26\linewidth]{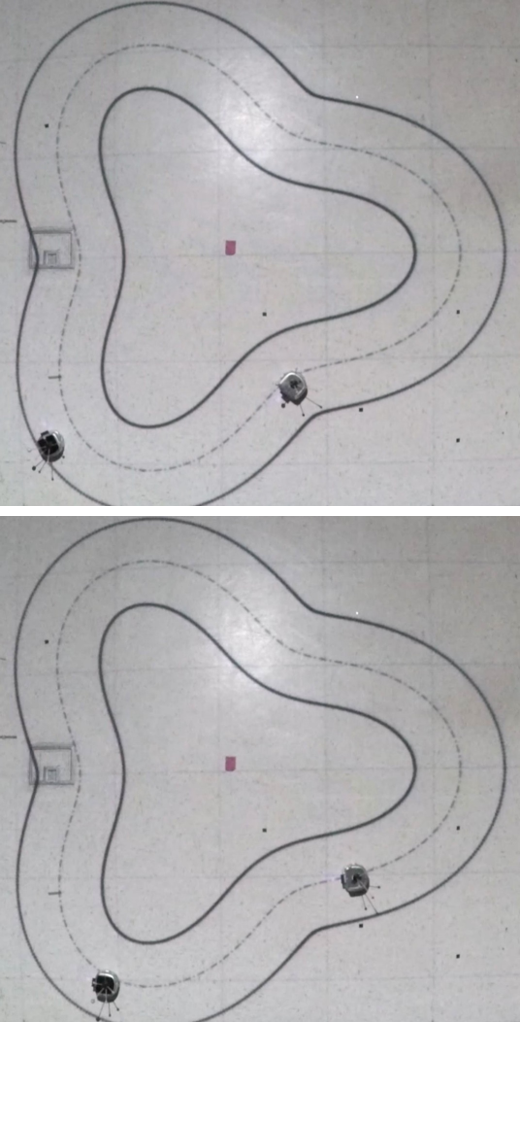} %
	    \label{fig:LK_ASR_expt}%
    }  
	\subfloat[][]{%
	    \includegraphics[width=0.74 \linewidth]{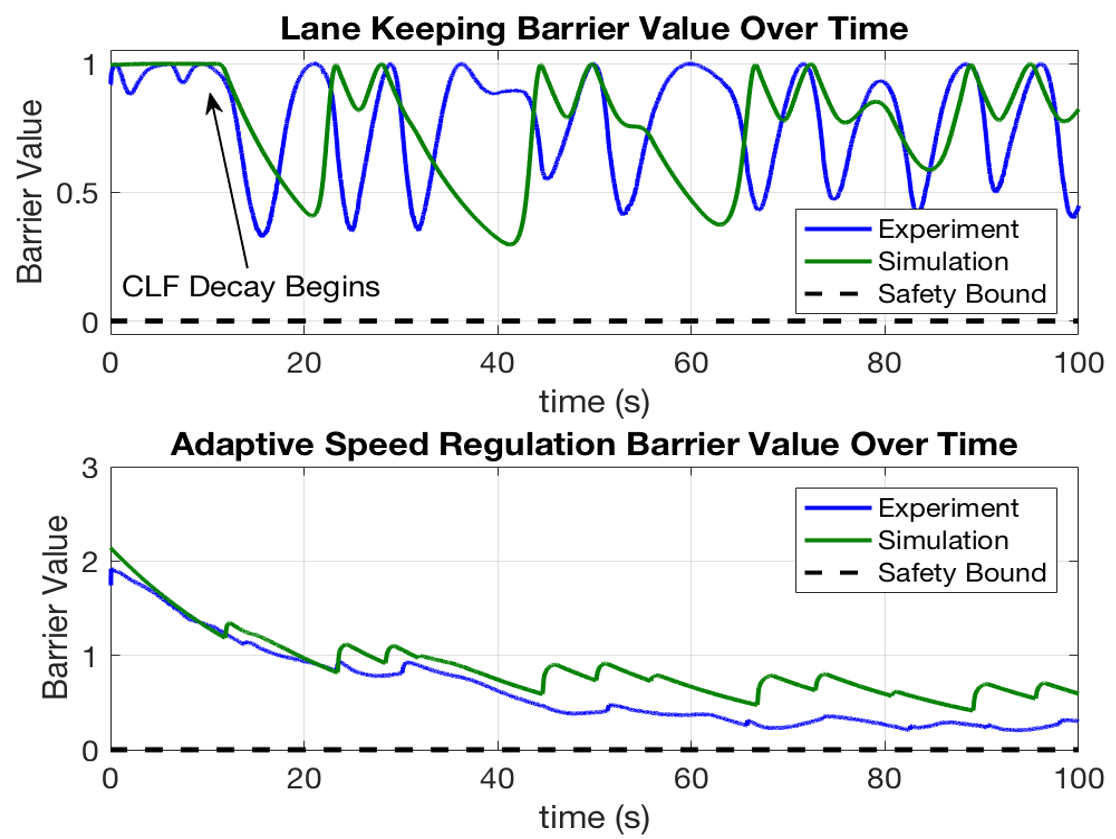}%
	    \label{fig:Khepera_CBF} 
    }
	\caption{Experimental demonstration of adaptive speed regulation and lane keeping for automotive systems.  
	(a) The robot is kept inside a lane due to the lane keeping CBF and follows another robot ahead by maintaining a fixed time-headway through the adaptive speed regulation CBF. 
	(b) Value of lane keeping, $h_{lk}$, and adaptive speed regulation, $h_{asr}$, CBFs for simulation and experiment.  Non-negativity of these values demonstrate enforcement of the constraints.  Video at \url{https://youtu.be/n_tTBq0TCYY}.}
	\label{fig:Khepera_robot_expts}
\end{figure}

Finally, the performance objectives such as driving the longitudinal velocity to a user-defined velocity ($v \to v_d$), creating a smoother path following ($\omega \to 0$), and following the desired path ($(x,y) \to R_d$) are specified through output functions that are regulated to zero through CLFs.  As earlier, the CLF and CBF conditions are unified into a single controller via \eqref{lect24:eqn:QPCLFCBF} given in Section~\ref{subsec:optbasedcontrol}.
Fig.~\ref{fig:LK_ASR_expt} shows experimental results on the Khepera robot where simultaneous enforcement of lane keeping and adaptive speed regulation safety constraints are enforced.  Fig.~\ref{fig:Khepera_CBF} illustrates the value of the CBFs in experiments and simulation.

\subsection{Dynamic Balancing on Segways}
\label{subsec:segway}
\begin{figure}[h!]
	\centering
	\includegraphics[width=1\linewidth]{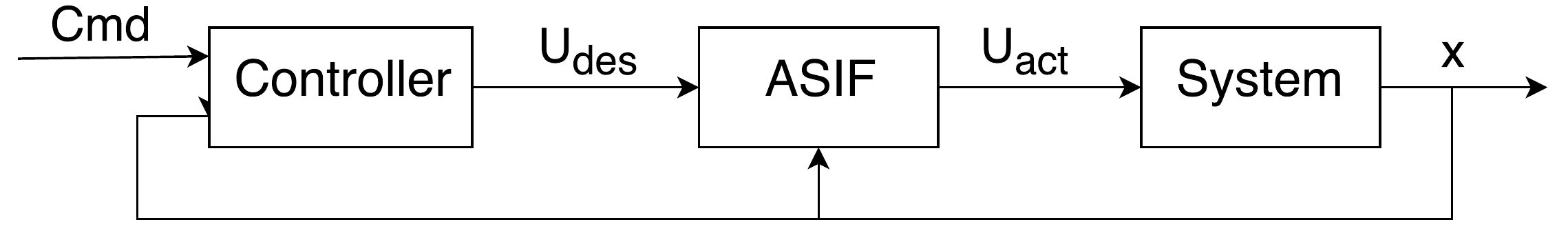}
	\caption{Figure illustrating the filtering of a desired control input through a safety filter, or Active Set Invariance Filter (ASIF).}
	\label{fig:ASIF}
\end{figure}

To demonstrate the application of control barrier functions as ``safety filters,'' we will consider their experimental realization on a Segway type robot, i.e., a two-wheeled inverted pendulum.  In particular, this subsection summarizes the results of \cite{gurriet2018towards} which provided the first experimental evaluation of CBFs on a robotic system that is not statically stable.  To realize these results, a Ninebot Segway was rebuilt, with only the original chassis and motors remaining---all of the electronics were customized to allow for the real-time control of the system via optimization based controllers.  The objective is to ensure ``safe'' operation of the Segway, defined in this case as the robot not tipping over, i.e., always staying upright.  Additionally, the goal is to achieve this safety condition even while using a nominal controller for the system (that may not be safe) and thus modifying the controller in a minimally invasive fashion so as to ensure safety.   The result will be a {\it safety filter}, or an {\it Active Set Invariance Filter (ASIF)} of the form illustrated in Fig. \ref{fig:ASIF}, where the nominal control input, $u_{\mathrm{des}}$, is filtered through a QP of the form \eqref{lectCBFpart1:eqn:mopt} to ensure safety in the system.  

The dynamics of the Segway can be written in the standard form given in \eqref{eqn:controlsys}, where in this case the input, $u$, is the voltage input into the motors and $x = (v,\phi,\dot{\phi})^T$, where $v$ is the forward velocity of the Segway, $\phi$ is the angle of the pendulum from upright, and $\dot{\phi}$ is the rate of change of this angle.  Correspondingly, there are input bounds on the system of the following form: $u \in [-15,15]\mathrm{V}$ (this input bounds will play a role in determining the CBF that will be implemented on hardware).  The safety constraint for the system is that the pendulum component of the robot stays upright, i.e., that the Segway does not tip over.  This can be captured by the condition that the angle of the pendulum, $\phi$, stays within a bounded region, in this case chosen to be $\phi \in [-\frac{\pi}{12},\frac{\pi}{12}] \mathrm{rad}$.  Finally, to ensure valid inputs, we also restrict the rate of change of the angle of the pendulum to be $\dot{\phi} \in [-2 \pi, 2 \pi] \mathrm{rad/s}$, and the forward velocity of the Segway to be $v \in [-5,5] \mathrm{m/s}$.  Finally, the nominal controller for the system, $u_{\mathrm{des}} = k(x)$, is chosen to be a standard PD controller that tracks a desired signal, i.e., an angle of the pendulum and velocity for the wheels. 

\begin{figure}[t!]
	\centering
	\subfloat[][]{%
	    \includegraphics[width=0.25\linewidth]{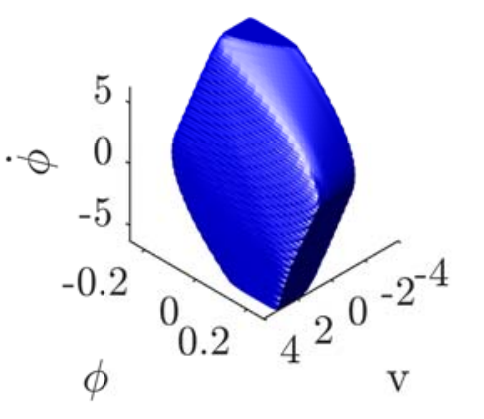}%
	    \label{fig:safeset}%
    } \hfill 
	\subfloat[][]{%
	    \includegraphics[width=0.75\linewidth]{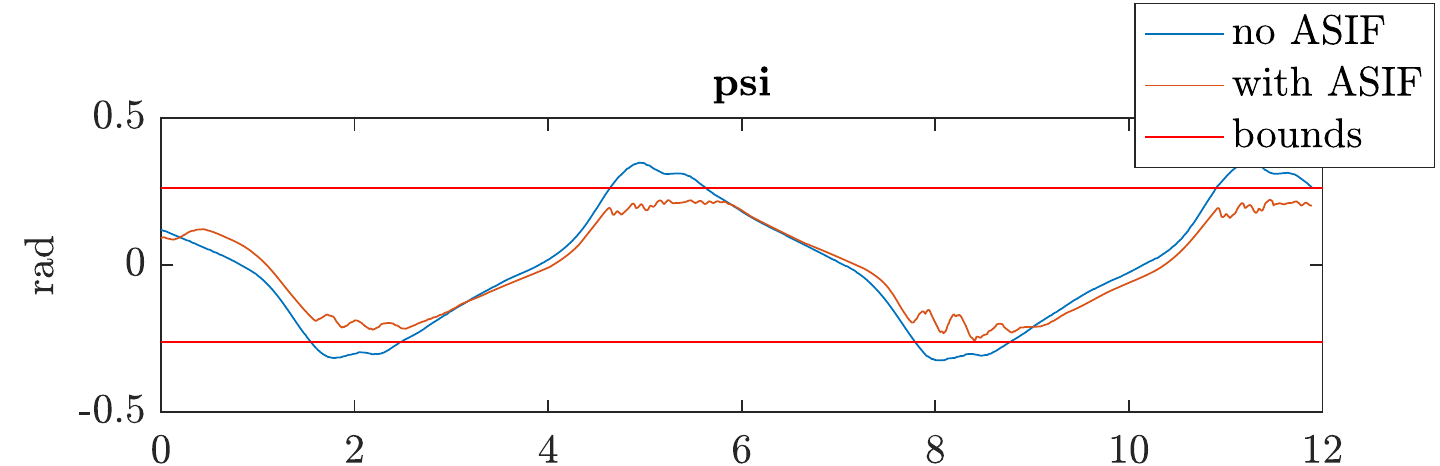}%
	    \label{fig:data_psi}%
    }\\
	\subfloat[][]{%
	    \includegraphics[width=1\linewidth]{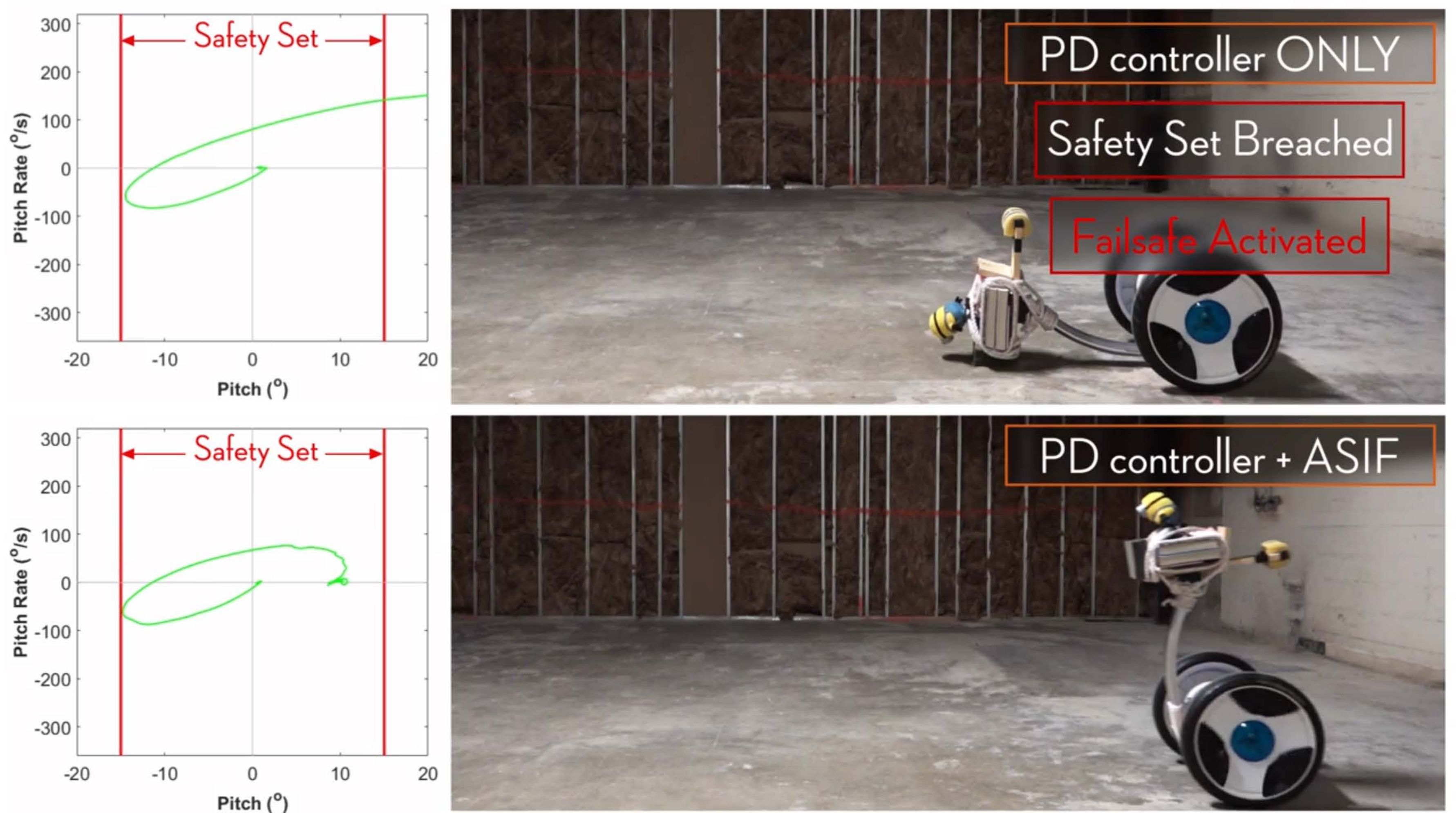}%
	    \label{fig:experimental}%
    }
	\caption{Experimental results for CBFs realized on a Segway robot to enforce safety defined as keeping the Segway upright, i.e., keeping the angle of the pedulum $\phi \in [-\frac{\pi}{12},\frac{\pi}{12}] \mathrm{rad}$, while satisfying additional phyiscal constraints. (a) The safe set $\C$ as calculated using Hamilton-Jacobi methods so that all physical realizability constraints are valid.  (b) Plots of the angle $\phi$ without and with the CBF (enforced via the ASIF). (b) Snapshot of experiment with and without the CBF implemented with an external disturbance (a kick)---in the case of no CBF, the Segway falls over. Experimental video: \url{https://youtu.be/RYXcGTo8Chg}.}
	\label{fig:segway}
\end{figure}

Since the safety constraint is to keep the Segway upright, i.e., keep $\phi \in [-\frac{\pi}{12},\frac{\pi}{12}] \mathrm{rad}$, one might be tempted to simply utilize two control barrier functions of the form: 
$$
h_1(\phi) = -\phi + \frac{\pi}{12}, \qquad h_2(\phi) = \phi - \frac{\pi}{12}.
$$
Yet, while these could be implemented via a CBF-QP to enforce these conditions, they will not enforce all of the additional constraints necessary to guarantee experimental implementation.  Therefore, the Hamilton-Jacobi method \cite{mitchell2005time} was utilized to determine the safe set $\C$ resulting by enforcing all the above-mentioned constraints. In particular, a reachability analysis was performed over a 75x75x75 grid of the state space with the edges of the grid at the state constraints given in the previous paragraph.  The resulting safe set can be seen in Fig. \ref{fig:safeset}.  A control barrier function can then be synthesized from this set---in this case, polynomial regression was used to create an analytic expression that can be used in the safety filter. 

The safety filter was implemented on hardware using the general framework indicated in Fig. \ref{fig:ASIF}.  In particular, the CLF-QP was solved onboard the hardware on a BeagleBone Black with an average computation time of 0.4 ms, with the resulting signal $u_{\mathrm{act}}$ passed to the motor controller.  To demonstrate the ability of the ASIF to enforce safety, the desired pendulum angle was passed to the system in the form of a sinusoidal signal with an amplitude exceeding the $\frac{\pi}{12}$ angle constraint.  Two experiments were then performed, one without and one with the ASIF, i.e., the CLF-QP active.  The results can be seen in Fig. \ref{fig:data_psi}, wherein the system remains safe only when the safety filter, implementing the CBF, is active.  Finally, to show the potential power of CBFs, a disturbance is added to the system in the form of a kick---the system is able to stay upright, and hence safe, with CBFs while the systems fails without them (illustrated in Fig. \ref{fig:experimental}).

%% file: sections/7_autonomy.tex
Another robotic application of CBFs involves the \textit{long duration autonomy} problem for multi-robot systems. This problem considers a team of robots deployed over long time scales which are asked to execute tasks (such as environmental monitoring, search and rescue, or precision agriculture) that require more than a  single charge of the battery of the robots. An effective control paradigm to use in this case is the constraint-based control \cite{egerstedt2018robot}, where \textit{survivability constraints}, i.e., conditions for the robots to remain operational over long temporal scales, can be enforced by means of CBFs and included in a single constrained optimization problem.

Consider a collection of $N$ mobile robots, whose dynamics are modeled by the following control affine system:
\begin{equation}
\nonumber
\dot x_i = f(x_i)+g(x_i)u_i,
\end{equation}
where $x_i\in\mathbb R^n$ and $u_i\in\mathbb R^m$, $i=1,\ldots,N$, are the state and the input of robot $i$, respectively, and $f$ and $g$ are locally Lipschitz. As the energy plays an important role in ensuring persistent operation, we augment the state $x_i$ by the energy $E_i$ stored in robot $i$'s battery obtaining: $\chi_i=[x_i^T, E_i]^T$. The energy dynamics are given by
\begin{equation}
\nonumber
\dot E_i = \hat f(\chi_i)+\hat g(\chi_i)u_i,
\end{equation}
where $\hat f$ and $\hat g$ are also assumed to be locally Lipschitz.
The dynamics of the augmented state $\chi_i$ are then:
\begin{equation}
\nonumber
\dot\chi_i = \begin{bmatrix}f(x_i)\\ \hat f(\chi_i)\end{bmatrix} + \begin{bmatrix}g(x_i)\\ \hat g(\chi_i)\end{bmatrix}u_i = F(\chi_i)+G(\chi_i)u_i.
\end{equation}
We assume the robot workspace is endowed with charging stations, interpreted as regions of the state space where robots can charge their batteries. Letting
\begin{equation}
\nonumber
p:x_i\in\mathbb R^n\mapsto p_i\in\mathbb R^d
\end{equation}
be a static mapping from robot $i$'s state to its position $p_i\in\mathbb R^d$, $d=2$ for ground robots or $d=3$ for aerial robots, we define
\begin{equation}
\nonumber
\rho_i:p_i\in\mathbb R^d\mapsto \rho_i(p_i)\in\mathbb R_{\ge0}
\end{equation}
as the function that evaluates the energy that robot $i$ requires to reach a charging station starting from position $p_i$.

We are now ready to encode the survivability constraints mentioned above. Following what has been done in \cite{notomista2018persistification}, \textit{survivability}, realized by ensuring that each robot never gets stranded away from a charging station, is encoded by ensuring that the following always holds:
\begin{equation}
\nonumber
h_{c,i}(\chi_i) = E_i - E_{min} - \rho_i(p(x_i)) \ge 0 \quad \forall i\in\{1,\ldots,N\},
\end{equation}
i.\,e. each robot always has enough energy to reach a charging station with a minimum desired amount of energy, $E_{min}$. Moreover, to prevent overcharging, we also want the following inequality to be always satisfied:
\begin{equation}
\nonumber
h_{o,i}(\chi_i) = E_{max} - E_i \ge 0.
\end{equation}
We can combine these two objectives by defining the logical \texttt{and} of these constraints, $h_{e,i} = h_{c,i} \land h_{o,i}$, as
\begin{equation}
\label{eq:and_energy}
h_{e,i}(\chi_i) = \min\{h_{c,i}(\chi_i),h_{o,i}(\chi_i)\},
\end{equation}
and enforcing differential constraints affine in the control variable $u_i$, which are analogous to \eqref{eqn:cbf:definition}, as shown in \cite{glotfelter2017nonsmooth}. 

Considering the environmental monitoring task, we reformulate the task itself using CBFs which can be then combined with the ones related to \textit{survivability} introduced above in order to implement persistent environmental monitoring \cite{acc2019ecologyarxiv}. Consider $N$ robots tasked with monitoring a compact and convex set $\Omega\subset\mathbb R^d$. We can define a measure of the coverage quality by \cite{cortes2004coverage}:
\begin{equation}
\label{eq:loccost}
J(x) = \sum_{i=1}^N \int_{\Omega_i}\|p(x_i)-q\|^2\phi(q)dq,
\end{equation}
where $x$ is the ensemble state of the robots, $\{\Omega_1,\ldots,\Omega_N\}$ is the Voronoi tessellation of the set $\Omega$, the value $\phi(q)\in\mathbb R,~\phi(q)\ge0~\forall q\in\Omega$, encodes the importance of the point $q$, and where the quality of the sensor coverage associated with the point $q$ decreases quadratically with the distance $\|p(x_i)-q\|$. The further away the point to monitor is, the worse the coverage is, and the higher the coverage cost $J$ is. Defining the barrier function related to the task as $h_{t}(\chi) = -J(x)$, where $\chi$ represents the ensemble compound state of the robots, containing $x_i$ and $E_i$ of each robot, we can express the constraint \eqref{eqn:cbf:definition} as
\begin{equation}
\label{eq:diff_constraint}
L_F h_{t}(\chi) + L_G h_{t}(\chi) u \geq - \alpha(h_{t}(\chi)).
\end{equation}
As shown in \cite{xu2015robustness}, the constraint \eqref{eq:diff_constraint} ensures that the zero superlevel set of the function $h_t(\chi)$ is asymptotically stable, with the effect of minimizing the coverage cost $J$ defined above \cite{acc2019ecologyarxiv}.

Additionally, safety, specifically intended as collision avoidance, can be guaranteed by ensuring that
\begin{equation}
\nonumber
h_{s}(\chi_i,\chi_j)=\|p(x_i)-p(x_j)\|^2-\Delta^2\ge0
\end{equation}
$\forall~i,j\in\{1,\ldots,N\},i\neq j$, where $\Delta>0$ is the safety distance to be maintained between any two robots, $i$ and $j$, located at positions $p(x_i)$ and $p(x_j)$. Similarly to what has been done to obtain \eqref{eq:and_energy}, we can define
\begin{equation}
\nonumber
h_i(\chi_i) = \min\left\{\min_{i}\left\{h_{e,i}(\chi_i)\right\},\min_{\substack{i,j\\i\neq j}}\left\{h_{s}(\chi_i,\chi_j)\right\}\right\},
\end{equation}
which combines energy and safety constraints, in order to formulate a differential constraint analogous to \eqref{eq:diff_constraint}.

Thus, each robot executes the input $u_i$ solution of the following QP:
\begin{align}
\label{eq:minuctrl}
\!\!\!\underset{u_1,\ldots,u_N,\delta}{\operatorname{min}}  ~& \sum_{i=1}^N \| u_i \|^2 + \kappa|\delta|^2  \\
\label{eq:surviv_constr}
\mathrm{s.t.}  ~& L_F h_i(\chi_i) + L_G h_i(\chi_i) u_i \geq - \alpha(h_i(\chi_i)),~\forall i\\
\nonumber
~& L_F h_{t}(\chi) + L_G h_{t}(\chi) u \geq - \alpha(h_{t}(\chi)) - \delta
\end{align}
where $\kappa>0$ is a weighting factor and the gradients involved in the computation of the Lie derivatives are intended as a particular class of generalized gradients (see \cite{glotfelter2017nonsmooth}). Note that introducing the relaxation variable $\delta$, as discussed in Section~\ref{sec:foundations}, allows us to trade the execution of the coverage task for safety and energy, i.\,e., survivability.

\begin{figure}
\centering
\subfloat[][]{\label{subfig:a}\includegraphics[trim={7cm 2cm 14cm 4cm}, clip,width=.24\textwidth]{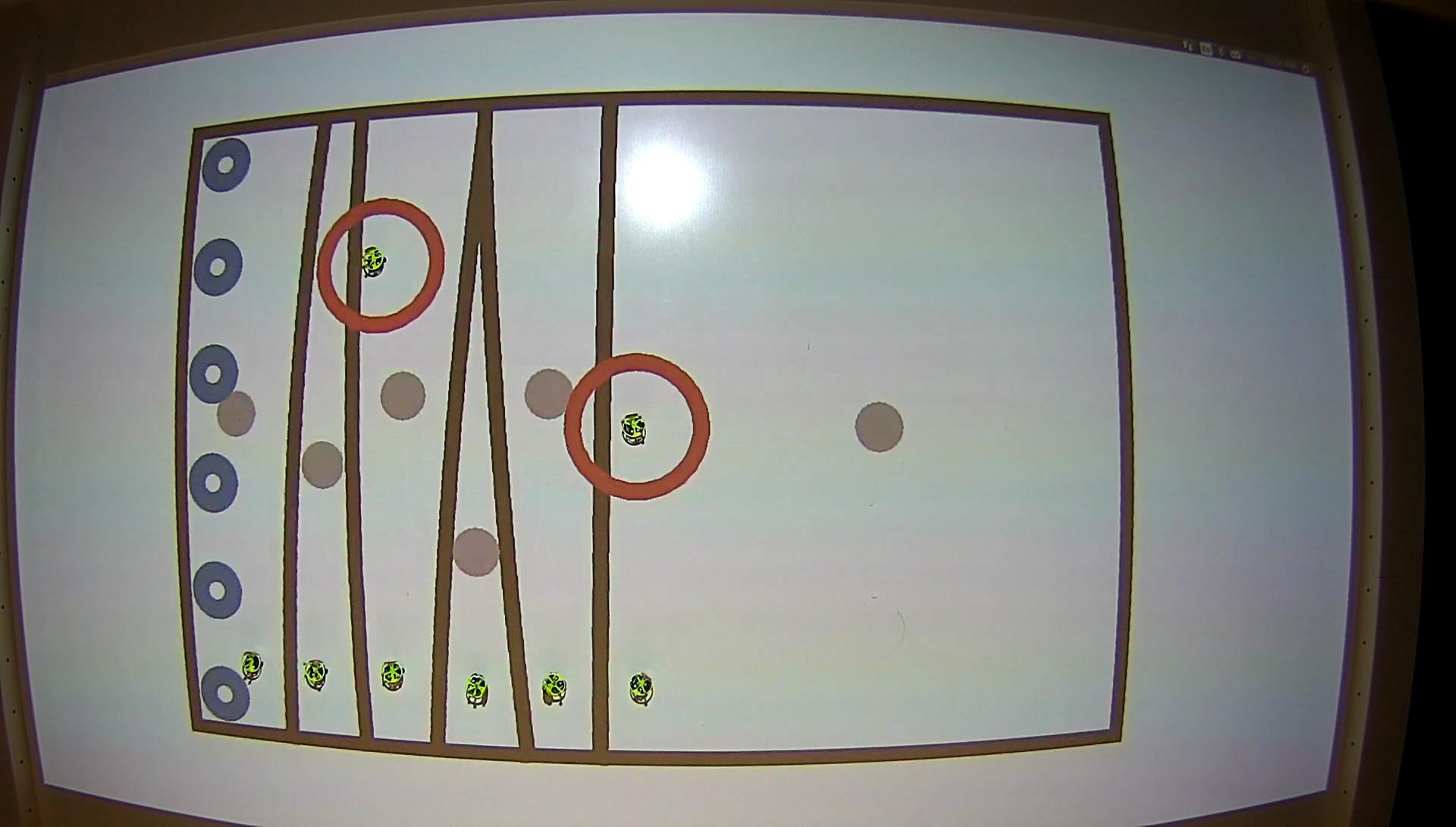}}\hfill
\subfloat[][]{\label{subfig:b}\includegraphics[trim={7cm 2cm 14cm 4cm}, clip,width=.24\textwidth]{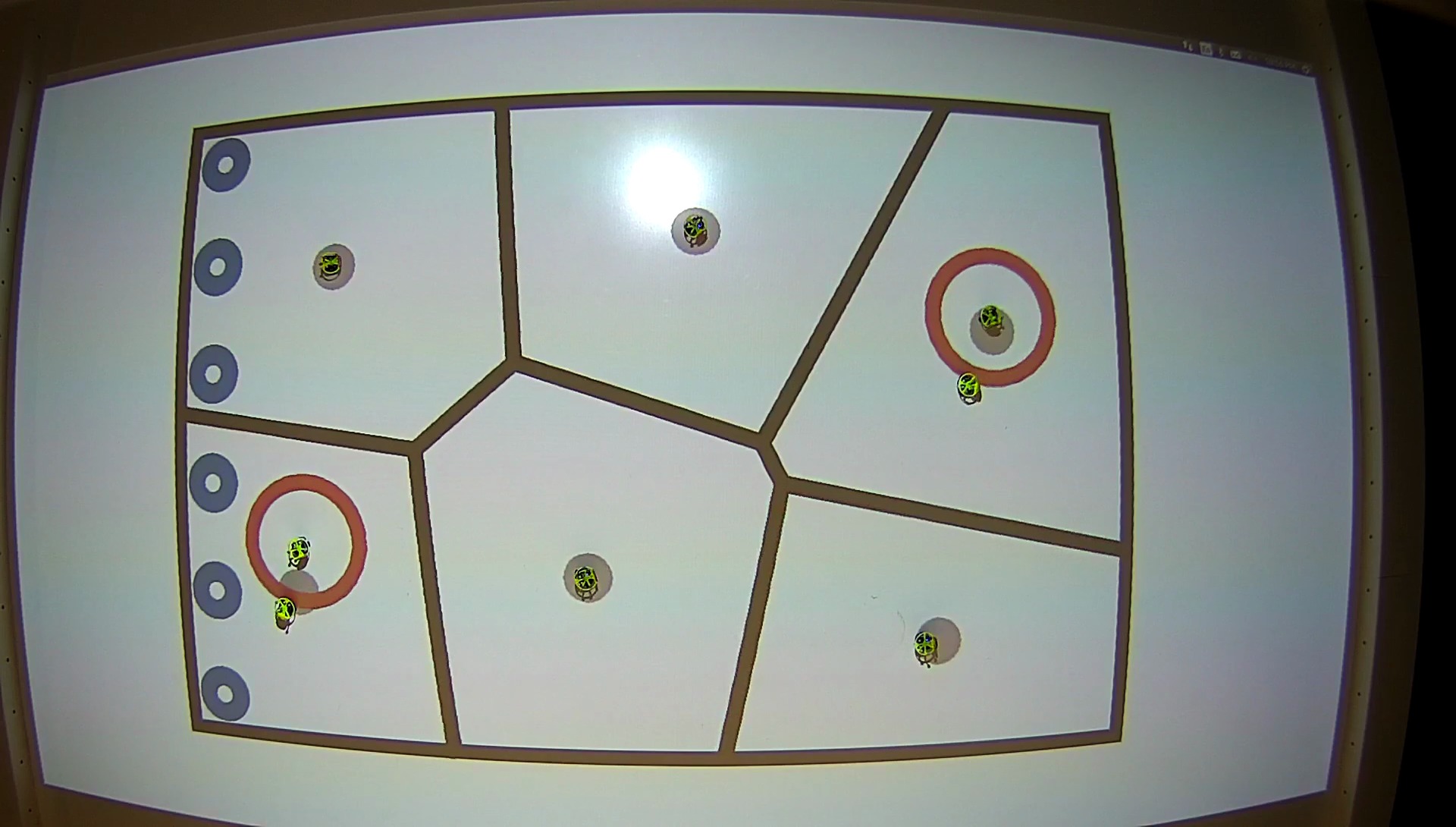}}\\
\subfloat[][]{\label{subfig:c}\includegraphics[trim={7cm 2cm 14cm 4cm}, clip,width=.24\textwidth]{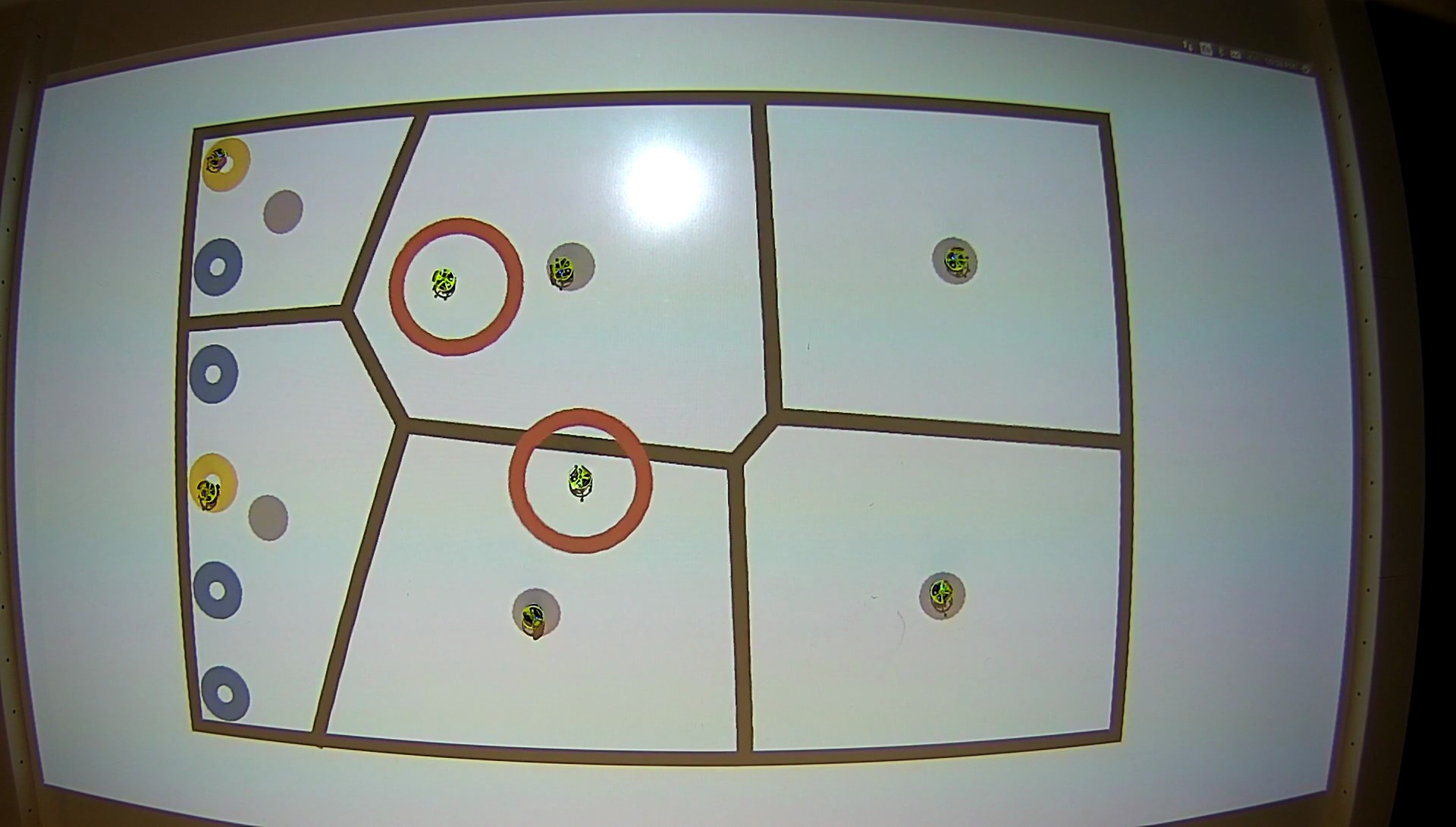}}\hfill
\subfloat[][]{\label{subfig:d}\includegraphics[trim={7cm 2cm 14cm 4cm}, clip,width=.24\textwidth]{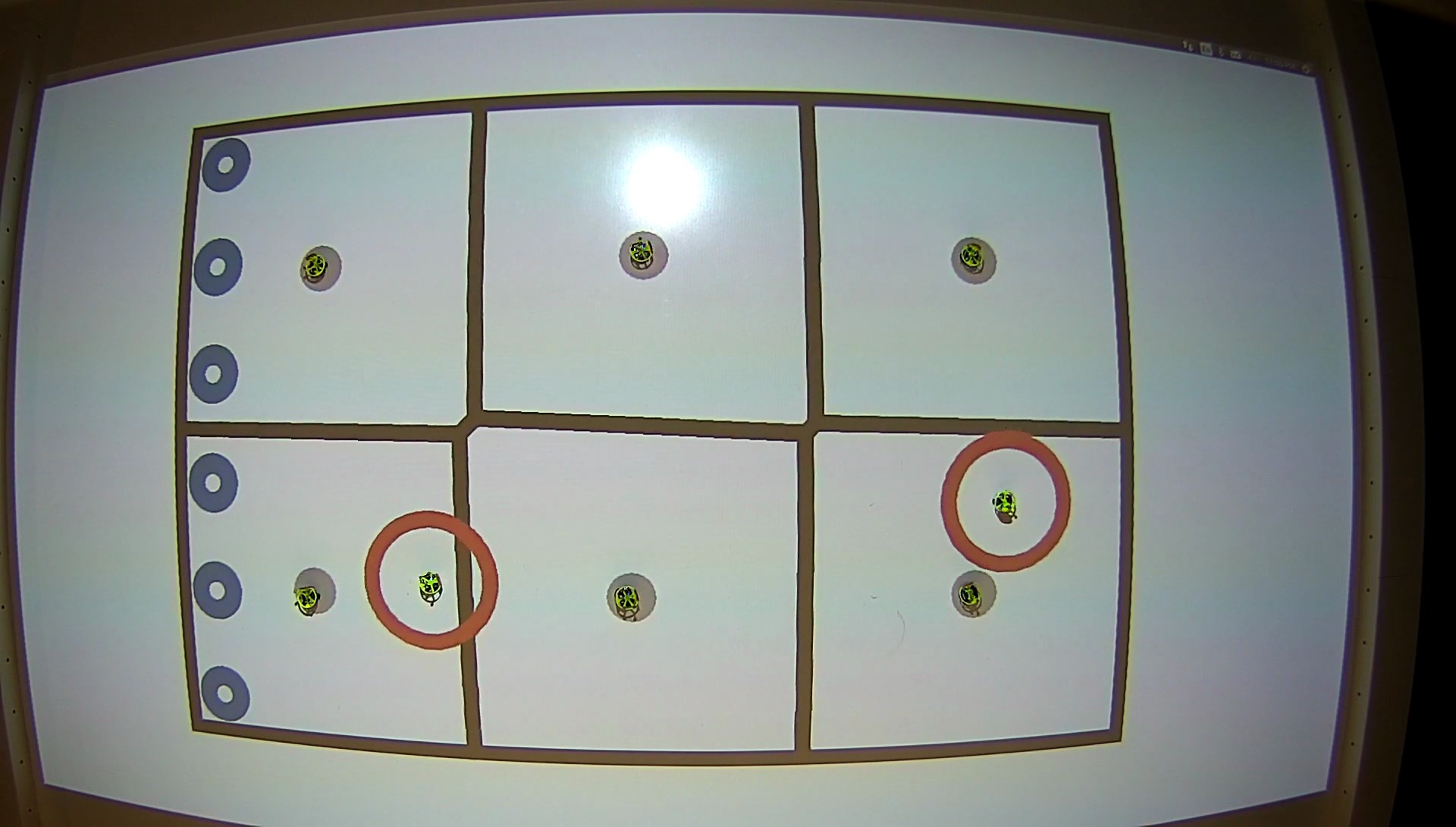}}
\caption{A team of six robots is tasked with monitoring a rectangular domain on the Robotarium, by performing coverage control. The boundary of the Voronoi partition is depicted using black thick lines. The robots are asked to perform this task over a time horizon which is much longer than their (simulated) battery life. Additionally, two more robots, circled in red, act as obstacles which have to be avoided by the remaining six robots. These execute the controller solution of \eqref{eq:minuctrl} to avoid the obstacles, go and recharge their batteries at the dedicated charging stations (blue circles on the left of the figures that turn yellow when the robots are charging), while always covering the given domain. A video of the experiments is available online at: \url{https://youtu.be/h-OTe4ieOrI}.
}
\label{fig:prioritizing}
\end{figure}

The persistent environmental monitoring strategy has been implemented on the Robotarium \cite{pickem2017robotarium}, where six ground mobile robots have been asked to monitor a given domain over a time horizon that is longer than their (simulated) battery life (see Fig.~\ref{fig:prioritizing}). The robots perform coverage control by minimizing the cost \eqref{eq:loccost} by enforcing the constraint \eqref{eq:diff_constraint}. Additionally, they have to avoid two obstacles moving in the environment (robots circled in red in Fig.~\ref{fig:prioritizing}) and never run out of energy. This is realized by means of the constraint \eqref{eq:surviv_constr}. Six charging stations (blue circles, which turn yellow when the robots are charging) allow the robots to recharge their battery. The charging stations are projected onto the testbed, together with the boundary of the Voronoi tessellation of the domain to cover. The execution of the controller solution of \eqref{eq:minuctrl} is summarized in Fig.~\ref{fig:prioritizing}.

%% file: sections/conclusions.tex
This paper presented a summary of recent results in safety-critical control based upon a novel form of control barrier functions.  The basis theoretic foundations of this formulation were reviewed, all with selected application domains.  Due to the recent activity in this domain, and the pressing need for safety in the context of autonomous systems, the authors imagine control barrier functions to become an essential component of modern control system design.  

%% file: main.bbl
% Generated by IEEEtran.bst, version: 1.14 (2015/08/26)
\begin{thebibliography}{10}
\providecommand{\url}[1]{#1}
\csname url@samestyle\endcsname
\providecommand{\newblock}{\relax}
\providecommand{\bibinfo}[2]{#2}
\providecommand{\BIBentrySTDinterwordspacing}{\spaceskip=0pt\relax}
\providecommand{\BIBentryALTinterwordstretchfactor}{4}
\providecommand{\BIBentryALTinterwordspacing}{\spaceskip=\fontdimen2\font plus
\BIBentryALTinterwordstretchfactor\fontdimen3\font minus
  \fontdimen4\font\relax}
\providecommand{\BIBforeignlanguage}[2]{{%
\expandafter\ifx\csname l@#1\endcsname\relax
\typeout{** WARNING: IEEEtran.bst: No hyphenation pattern has been}%
\typeout{** loaded for the language `#1'. Using the pattern for}%
\typeout{** the default language instead.}%
\else
\language=\csname l@#1\endcsname
\fi
#2}}
\providecommand{\BIBdecl}{\relax}
\BIBdecl

\bibitem{LL77}
L.~Lamport, ``Proving the correctness of multiprocess programs,'' \emph{IEEE
  Transactions on Control Engineering}, vol.~3, no.~2, pp. 125--143, 1977.

\bibitem{LL84}
------, ``Basic concepts,'' in \emph{Advanced Course on Distributed
  Systems--Methods and Tools for Specification}, ser. Lecture Notes in Computer
  Science, vol. 190.\hskip 1em plus 0.5em minus 0.4em\relax Springer, 1984.

\bibitem{HS85}
B.~Alpern and F.~B. Schneider, ``Defining liveness,'' \emph{Information
  Processing Letters}, pp. 181--185, 1985.

\bibitem{nagumo1942lage}
M.~Nagumo, ``{\"U}ber die lage der integralkurven gew{\"o}hnlicher
  differentialgleichungen,'' \emph{Proceedings of the Physico-Mathematical
  Society of Japan. 3rd Series}, vol.~24, pp. 551--559, 1942.

\bibitem{blanchini1999set}
F.~Blanchini, ``Set invariance in control,'' \emph{Automatica}, vol.~35,
  no.~11, pp. 1747--1767, 1999.

\bibitem{abraham2012manifolds}
R.~Abraham, J.~E. Marsden, and T.~Ratiu, \emph{Manifolds, tensor analysis, and
  applications}.\hskip 1em plus 0.5em minus 0.4em\relax Springer Science \&
  Business Media, 2012, vol.~75.

\bibitem{bony1969principe}
J.-M. Bony, ``Principe du maximum, in{\'e}galit{\'e} de harnack et unicit{\'e}
  du probleme de cauchy pour les op{\'e}rateurs elliptiques
  d{\'e}g{\'e}n{\'e}r{\'e}s,'' \emph{Ann. Inst. Fourier (Grenoble)}, vol.~19,
  no.~1, pp. 277--304, 1969.

\bibitem{brezis1970characterization}
H.~Brezis, ``On a characterization of flow-invariant sets,''
  \emph{Communications on Pure and Applied Mathematics}, vol.~23, no.~2, pp.
  261--263, 1970.

\bibitem{prajna2004safety}
S.~Prajna and A.~Jadbabaie, ``Safety verification of hybrid systems using
  barrier certificates,'' in \emph{International Workshop on Hybrid Systems:
  Computation and Control}.\hskip 1em plus 0.5em minus 0.4em\relax Springer,
  2004, pp. 477--492.

\bibitem{prajna2006barrier}
S.~Prajna, ``Barrier certificates for nonlinear model validation,''
  \emph{Automatica}, vol.~42, no.~1, pp. 117--126, 2006.

\bibitem{prajna2005necessity}
S.~Prajna and A.~Rantzer, ``On the necessity of barrier certificates,''
  \emph{IFAC Proceedings Volumes}, vol.~38, no.~1, pp. 526--531, 2005.

\bibitem{prajna2007framework}
S.~Prajna, A.~Jadbabaie, and G.~J. Pappas, ``A framework for worst-case and
  stochastic safety verification using barrier certificates,'' \emph{IEEE
  Transactions on Automatic Control}, vol.~52, no.~8, pp. 1415--1428, 2007.

\bibitem{tee2009barrier}
K.~P. Tee, S.~S. Ge, and E.~H. Tay, ``Barrier lyapunov functions for the
  control of output-constrained nonlinear systems,'' \emph{Automatica},
  vol.~45, no.~4, pp. 918--927, 2009.

\bibitem{aubin2009viability}
J.-P. Aubin, \emph{Viability theory}.\hskip 1em plus 0.5em minus 0.4em\relax
  Springer Science \& Business Media, 2009.

\bibitem{aubin1990survey}
------, ``A survey of viability theory,'' \emph{SIAM Journal on Control and
  Optimization}, vol.~28, no.~4, pp. 749--788, 1990.

\bibitem{aubin2011viability}
J.-P. Aubin, A.~M. Bayen, and P.~Saint-Pierre, \emph{Viability theory: new
  directions}.\hskip 1em plus 0.5em minus 0.4em\relax Springer Science \&
  Business Media, 2011.

\bibitem{wieland2007constructive}
P.~Wieland and F.~Allg{\"o}wer, ``Constructive safety using control barrier
  functions,'' \emph{IFAC Proceedings Volumes}, vol.~40, no.~12, pp. 462--467,
  2007.

\bibitem{romdlony2014uniting}
M.~Z. Romdlony and B.~Jayawardhana, ``Uniting control lyapunov and control
  barrier functions,'' in \emph{Decision and Control (CDC), 2014 IEEE 53rd
  Annual Conference on}.\hskip 1em plus 0.5em minus 0.4em\relax IEEE, 2014, pp.
  2293--2298.

\bibitem{romdlony2016stabilization}
------, ``Stabilization with guaranteed safety using control lyapunov--barrier
  function,'' \emph{Automatica}, vol.~66, pp. 39--47, 2016.

\bibitem{ames2014control}
A.~D. Ames, J.~W. Grizzle, and P.~Tabuada, ``Control barrier function based
  quadratic programs with application to adaptive cruise control,'' in
  \emph{Decision and Control (CDC), 2014 IEEE 53rd Annual Conference on}.\hskip
  1em plus 0.5em minus 0.4em\relax IEEE, 2014, pp. 6271--6278.

\bibitem{ames2017cbf}
A.~D. Ames, X.~Xu, J.~W. Grizzle, and P.~Tabuada, ``Control barrier function
  based quadratic programs for safety critical systems,'' \emph{IEEE
  Transactions on Automatic Control}, vol.~62, no.~8, pp. 3861--3876, 2017.

\bibitem{xu2015robustness}
X.~Xu, P.~Tabuada, J.~W. Grizzle, and A.~D. Ames, ``Robustness of control
  barrier functions for safety critical control,'' \emph{IFAC-PapersOnLine},
  vol.~48, no.~27, pp. 54--61, 2015.

\bibitem{Xu2017_ccta_lk_asr}
X.~Xu, T.~Waters, D.~Pickem, P.~Glotfelter, M.~Egerstedt, P.~Tabuada, J.~W.
  Grizzle, and A.~D. Ames, ``Realizing simultaneous lane keeping and adaptive
  speed regulation on accessible mobile robot testbeds,'' in \emph{IEEE
  Conference on Control Technology and Applications}, Mauna Lani, HI, August
  2017, pp. 1769--1775.

\bibitem{Xu:2018tk}
X.~Xu, J.~W. Grizzle, P.~Tabuada, and A.~D. Ames, ``Correctness guarantees for
  the composition of lane keeping and adaptive cruise control,'' \emph{IEEE
  Transactions on Automation Science and Engineering}, vol.~15, no.~3, pp.
  1216--1229, 2018.

\bibitem{borrmann2015control}
U.~Borrmann, L.~Wang, A.~D. Ames, and M.~B. Egerstedt, ``Control barrier
  certificates for safe swarm behavior.''\hskip 1em plus 0.5em minus
  0.4em\relax Georgia Institute of Technology, 2015.

\bibitem{wang2017safety}
L.~Wang, A.~D. Ames, and M.~Egerstedt, ``Safety barrier certificates for
  collisions-free multirobot systems,'' \emph{IEEE Transactions on Robotics},
  vol.~33, no.~3, pp. 661--674, 2017.

\bibitem{pickem2017robotarium}
D.~Pickem, P.~Glotfelter, L.~Wang, M.~Mote, A.~Ames, E.~Feron, and
  M.~Egerstedt, ``The robotarium: A remotely accessible swarm robotics research
  testbed,'' in \emph{Robotics and Automation (ICRA), 2017 IEEE International
  Conference on}.\hskip 1em plus 0.5em minus 0.4em\relax IEEE, 2017, pp.
  1699--1706.

\bibitem{wu2016safety}
G.~Wu and K.~Sreenath, ``Safety-critical control of a planar quadrotor,'' in
  \emph{American Control Conference (ACC), 2016}.\hskip 1em plus 0.5em minus
  0.4em\relax IEEE, 2016, pp. 2252--2258.

\bibitem{wang2017safe}
L.~Wang, A.~D. Ames, and M.~Egerstedt, ``Safe certificate-based maneuvers for
  teams of quadrotors using differential flatness,'' in \emph{IEEE
  International Conference on Robotics and Automation (ICRA)}, 2017.

\bibitem{CBF:Ames:CBFBiped:ACC15}
S.-C. Hsu, X.~Xu, and A.~D. Ames, ``Control barrier function based quadratic
  programs with application to bipedal robotic walking,'' in \emph{American
  Control Conference}, 2015.

\bibitem{CDC2016_3DWalking_SteppingStones}
Q.~Nguyen, A.~Hereid, J.~W. Grizzle, A.~D. Ames, and K.~Sreenath, ``3d dynamic
  walking on stepping stones with control barrier functions,'' in \emph{IEEE
  International Conference on Decision and Control (CDC)}, Las Vegas, NV,
  December 2016, pp. 827--834.

\bibitem{gurriet2018towards}
T.~Gurriet, A.~Singletary, J.~Reher, L.~Ciarletta, E.~Feron, and A.~D. Ames,
  ``Towards a framework for realizable safety critical control through active
  set invariance,'' in \emph{Proceedings of the 9th ACM/IEEE International
  Conference on Cyber-Physical Systems}.\hskip 1em plus 0.5em minus 0.4em\relax
  IEEE Press, 2018, pp. 98--106.

\bibitem{khalil2015nonlinear}
H.~K. Khalil, \emph{Nonlinear control}.\hskip 1em plus 0.5em minus 0.4em\relax
  Pearson New York, 2015.

\bibitem{Sontag:firstCLF}
E.~Sontag, ``A {L}yapunov-like stabilization of asymptotic controllability,''
  \emph{SIAM Journal of Control and Optimization}, vol.~21, no.~3, pp.
  462--471, 1983.

\bibitem{Sontag:universal}
------, ``A 'universal' contruction of {A}rtstein's theorem on nonlinear
  stabilization,'' \emph{Systems \& Control Letters}, vol.~13, pp. 117--123,
  1989.

\bibitem{artstein1983stabilization}
Z.~Artstein, ``Stabilization with relaxed controls,'' \emph{Nonlinear Analysis:
  Theory, Methods \& Applications}, vol.~7, no.~11, pp. 1163--1173, 1983.

\bibitem{ames2014rapidly}
A.~D. Ames, K.~Galloway, K.~Sreenath, and J.~W. Grizzle, ``Rapidly
  exponentially stabilizing control lyapunov functions and hybrid zero
  dynamics,'' \emph{IEEE Transactions on Automatic Control}, vol.~59, no.~4,
  pp. 876--891, 2014.

\bibitem{Tee}
K.~P. Tee, S.~S. Ge, and E.~H. Tay, ``Barrier {L}yapunov functions for the
  control of output-constrained nonlinear systems,'' \emph{Automatica},
  vol.~45, no.~4, pp. 918 -- 927, 2009.

\bibitem{boyd2004convex}
S.~Boyd and L.~Vandenberghe, \emph{Convex optimization}.\hskip 1em plus 0.5em
  minus 0.4em\relax Cambridge university press, 2004.

\bibitem{freeman2008robust}
R.~Freeman and P.~V. Kokotovic, \emph{Robust nonlinear control design:
  state-space and Lyapunov techniques}.\hskip 1em plus 0.5em minus 0.4em\relax
  Springer Science \& Business Media, 2008.

\bibitem{ames2013towards}
A.~D. Ames and M.~Powell, ``Towards the unification of locomotion and
  manipulation through control lyapunov functions and quadratic programs,'' in
  \emph{Control of Cyber-Physical Systems}.\hskip 1em plus 0.5em minus
  0.4em\relax Springer, 2013, pp. 219--240.

\bibitem{galloway2015torque}
K.~Galloway, K.~Sreenath, A.~D. Ames, and J.~W. Grizzle, ``Torque saturation in
  bipedal robotic walking through control lyapunov function-based quadratic
  programs,'' \emph{IEEE Access}, vol.~3, pp. 323--332, 2015.

\bibitem{Squires:2018nr}
E.~Squires, P.~Pierpaoli, and M.~Egerstedt, ``Constructive barrier certificates
  with applications to fixed-wing aircraft collision avoidance,'' in \emph{2018
  IEEE Conference on Control Technology and Applications (CCTA)}, Aug 2018, pp.
  1656--1661.

\bibitem{Parrilo:2001uq}
P.~A. Parrilo, ``Semidefinite programming relaxations for semialgebraic
  problems,'' \emph{Mathematical Programming Ser. B}, vol.~96, no.~2, pp.
  293--320, 2003.

\bibitem{wang2018permissive}
L.~Wang, D.~Han, and M.~Egerstedt, ``Permissive barrier certificates for safe
  stabilization using sum-of-squares,'' in \emph{American Control Conference},
  2018, pp. 585--590.

\bibitem{sostools}
S.~Prajna, A.~Papachristodoulou, P.~Seiler, and P.~A. Parrilo,
  \emph{{SOSTOOLS}: Sum of squares optimization toolbox for {MATLAB}},
  \texttt{http://www.cds.caltech.edu/sostools}, 2018.

\bibitem{ACC2015_Geometric_CBFCLF}
G.~Wu and K.~Sreenath, ``Safety-critical and constrained geometric control
  synthesis using control lyapunov and control barrier functions for systems
  evolving on manifolds,'' in \emph{American Control Conference (ACC)},
  Chicago, IL, July 2015, pp. 2038--2044.

\bibitem{ACC2016_Exponential_CBF}
Q.~Nguyen and K.~Sreenath, ``Exponential control barrier functions for
  enforcing high relative-degree safety-critical constraints,'' in
  \emph{American Control Conference (ACC)}, Boston, MA, July 2016, pp.
  322--328.

\bibitem{ADHS2015_FootstepCBF}
------, ``Safety-critical control for dynamical bipedal walking with precise
  footstep placement,'' in \emph{IFAC Analysis and Design of Hybrid Systems
  (ADHS)}, Atlanta, GA, October 2015.

\bibitem{WGCCM07}
E.~R. Westervelt, J.~W. Grizzle, C.~Chevallereau, J.~Choi, and B.~Morris,
  \emph{Feedback Control of Dynamic Bipedal Robot Locomotion}, ser. Control and
  Automation, Boca Raton, FL, June 2007.

\bibitem{mitchell2005time}
I.~M. Mitchell, A.~M. Bayen, and C.~J. Tomlin, ``A time-dependent
  hamilton-jacobi formulation of reachable sets for continuous dynamic games,''
  \emph{IEEE Transactions on automatic control}, vol.~50, no.~7, pp. 947--957,
  2005.

\bibitem{egerstedt2018robot}
M.~Egerstedt, J.~N. Pauli, G.~Notomista, and S.~Hutchinson, ``Robot ecology:
  Constraint-based control design for long duration autonomy,'' \emph{Annual
  Reviews in Control}, 2018.

\bibitem{notomista2018persistification}
G.~Notomista, S.~F. Ruf, and M.~Egerstedt, ``Persistification of robotic tasks
  using control barrier functions,'' \emph{IEEE Robotics and Automation
  Letters}, vol.~3, no.~2, pp. 758--763, 2018.

\bibitem{glotfelter2017nonsmooth}
P.~Glotfelter, J.~Cort{\'e}s, and M.~Egerstedt, ``Nonsmooth barrier functions
  with applications to multi-robot systems,'' \emph{IEEE control systems
  letters}, vol.~1, no.~2, pp. 310--315, 2017.

\bibitem{acc2019ecologyarxiv}
G.~Notomista and M.~Egerstedt, ``Constraint-driven coordinated control of
  multi-robot systems,'' \emph{arXiv preprint arXiv:1811.02465}, 2018.

\bibitem{cortes2004coverage}
J.~Cortes, S.~Martinez, T.~Karatas, and F.~Bullo, ``Coverage control for mobile
  sensing networks,'' \emph{IEEE Transactions on robotics and Automation},
  vol.~20, no.~2, pp. 243--255, 2004.

\end{thebibliography}
